\documentclass[a4paper,fleqn]{cas-dc}
\usepackage{graphicx,subcaption}
\usepackage{multicol}
\usepackage{layout}
\usepackage{amsfonts}
\usepackage{amsmath}
\usepackage{amssymb}
\usepackage{textcomp}
\usepackage{caption}
\usepackage[normalem]{ulem}
\usepackage[utf8]{inputenc}
\usepackage{natbib} 
\usepackage{rotating}
\usepackage{lscape}
\usepackage[justification=centering]{caption}
\usepackage{aas_macros}
\usepackage{threeparttable}
\usepackage{epstopdf}
\usepackage{comment}
\usepackage{soul}
\newcommand{\gm}{$\gamma$}

\begin{document}
\UseRawInputEncoding
\let\WriteBookmarks\relax
\def\floatpagepagefraction{1}
\def\textpagefraction{.001}
\newcommand{\cdr}[1]{{\bf{\color{red} #1}}}
\newcommand{\cdrtwo}[2]{{\st{#1}}  {{\bf \color{blue} #2}}} 
\shorttitle{VHE emission from FSRQs}
\shortauthors{Naseef et al.}

\title[mode = title]{\textit{Fermi}-Large Area Telescope Detection of Very High Energy (>100 GeV) Emission from Compton-Dominated Blazars}

\author[1,2]{P N {Naseef Mohammed}}[orcid=0000-0003-0545-5998]

\cormark[1]


\ead{naseefmhd@farookcollege.ac.in}



\affiliation[1]{organization={Farook College},
            city={Calicut},
            postcode={673632}, 
            state={Kerala},
            country={India}} 
\affiliation[2]{organization={University of Calicut},
            city={Malappuram},
            postcode={673635}, 
            state={Kerala},
            country={India}}
           
\author[3]{Vaidehi S. Paliya}[]
\ead{vaidehi.s.paliya@gmail.com}

\author[2]{C D Ravikumar}[]
\ead{cdr@uoc.ac.in}

\affiliation[3]{organization={Inter-University Centre for Astronomy and Astrophysics (IUCAA), SPPU Campus},
            city={Pune},
            postcode={411007}, 
            state={Maharashtra},
            country={India}}

\cortext[cor1]{Corresponding author}


\nonumnote{}

\begin{abstract}
The observation of broad emission lines in the optical spectra of flat-spectrum radio quasars (FSRQs) suggests radiatively efficient accretion powering these objects. In such broad emission line blazars, the intense broad-line region (BLR) radiation can provide seed photons for inverse Compton scattering, leading to a Compton-dominated spectral energy distribution. Interestingly, the same BLR photon field can also absorb very high-energy (VHE; E$>$100 GeV) $\gamma$-ray radiation, thus explaining the paucity of VHE-detected FSRQs. Here we report the results of a systematic search to identify VHE-emitting sources in a sample of 626 Compton-dominated blazars (Compton dominance $>$1), using $\sim$17.5 years of Fermi-Large Area Telescope observations. We identified 14 blazars at $\gtrsim4\sigma$ confidence level, including 4 sources detected in the VHE band at high significance ($>5\sigma$) for the first time. We also found 21 objects from which at least one VHE photon was detected, thus substantially expanding the known VHE FSRQ population. Investigating the temporal coincidence of the VHE photons with the $\gamma$-ray activity, we noticed the VHE emission to be detected during flaring as well as low jet activity epochs. By estimating the optical depth for the $\gamma\gamma$ absorption due to the BLR photon field, we constrained the VHE-emitting region to be located outside BLR ($>$1.1-1.4$\times$ BLR radius). We conclude that multi-wavelength followup observations of these enigmatic VHE-detected broad line blazars will permit us to constrain the radiative processes responsible for the GeV$-$TeV emission, and will set the benchmark for their observations with the upcoming Cherenkov Telescope Array Observatory.

\end{abstract}

\begin{keywords}
Galaxies: active \sep Methods: data analysis \sep FSRQ \sep gamma rays \sep VHE
\end{keywords}

\maketitle
\section{Introduction}
Blazars are a subclass of radio-loud active galactic nuclei (AGN) distinguished by relativistic jets oriented close to the line of sight to the observer \citep{1995PASP..107..803U,2019ARA&A..57..467B}. This alignment leads to strong Doppler boosting of the jet emission, resulting in brightness enhancement and large amplitude flux variations across the entire electromagnetic spectrum. Their broadband spectral energy distributions (SEDs) exhibit a characteristic double-humped structure, indicative of two distinct non-thermal emission processes. The low-energy hump, extending from the far-infrared to the soft X-ray regime, is attributed to synchrotron radiation from relativistic electrons in the jet. The high-energy component, extending from hard X-rays to $\gamma$-rays and occasionally into the TeV range, is generally ascribed to Inverse Compton scattering of relativistic electrons by photons internal or external to the jet or to hadronic emission processes \citep{2000AIPC..515...41R,2009ApJ...692...32D,2010ApJ...716...30A}

Gamma-ray astronomy, particularly in the Very High-Energy (VHE; $E \ge 100\,\mathrm{GeV}$) band, provides unique insights into extreme astrophysical environments where particles are accelerated and radiative processes operate near theoretical limits. Identifying potential VHE $\gamma$-ray candidates has become increasingly important with the rise of multimessenger astrophysics and the improved sensitivity expected from the next generation of VHE observatories for a more targeted sky survey. However, current VHE astronomy is constrained by a limited number of known TeV-emitting sources, as reflected in the relatively sparse entries of the TeVCat catalog and by observational challenges inherent to imaging atmospheric Cherenkov telescopes (IACTs) such as Major Atmospheric Gamma Imaging Cherenkov telescope  \citep[MAGIC;][]{2016APh....72...76A}, Very Energetic Radiation Imaging Telescope Array System \citep[VERITAS;][]{2006APh....25..391H} and High Energy Stereoscopic System \citep[H.E.S.S.][]{2004NewAR..48..331H}.

Ground-based IACTs are designed to detect $\gamma$-rays extending up to energies of tens of TeV, with high sensitivity and angular resolution. However, they have limitations due to restricted sky coverage, dependence on weather conditions, clear night skies, and source visibility, as observations are limited to specific times and locations. The Fermi Large Area Telescope (LAT) has revolutionized our understanding of the high-energy (HE; $100\,\mathrm{MeV} \text{--} 100\,\mathrm{GeV}$) $\gamma$-ray sky since August 2008, particularly in probing emissions from blazars \citep[cf.][]{4lac_22}. Its capacity to conduct continuous all-sky survey, from 20 MeV to over 2 TeV, has yielded several catalogs, with the latest being the Fermi-LAT 16 years source list (FL16Y) comprising of  7220 sources up to 1 TeV \citep{2026arXiv260222148B}, $\sim 56\%$ of which are AGN including blazars (835 flat spectrum radio quasars, 1825 BL Lacs, and 1516 blazars of uncertain type), narrow-line Seyfert 1 galaxies, and misaligned systems such as radio galaxies \citep{2020ApJS..247...33A,4fgl_dr4,2024ApJ...976..120P,2026arXiv260222148B}. Despite reduced sensitivity above $\sim$10 GeV, Fermi-LAT's large field of view and continuous exposure effectively mitigate background contamination, enabling identification of promising VHE blazar candidates for follow-up with ground-based instruments and provide spectral overlap with the VHE regime \citep[e.g.,][]{2025ApJ...991L...8P,2025arXiv250608497N,2026JHEAp..4900454T,2026ApJS..282...35A}.

Among the blazar population, BL Lac objects are the most numerous class of VHE-emitting sources \citep[cf.][]{2026ApJS..282...35A}. This can be explained considering these objects have a weak accretion activity, thus making the radiative environment surrounding the jet sparse. Therefore, jet electrons can accelerate to very high energies and emit GeV-TeV radiation via synchrotron self-Compton radiation \citep[][]{2010MNRAS.401.1570T,2018MNRAS.477.4257C}. On the other hand, the detection of broad emission lines from flat spectrum radio quasars (FSRQs) suggests them to be powered by a luminous accretion disk, which can photoionize both broad and narrow line region clouds and illuminate the dusty torus, thus providing a rich environment for inverse Compton scattering. Accordingly, in FSRQs, jet electrons quickly lose energy before attaining very high energies, thereby making them brighter at MeV-GeV energies and exhibiting a Compton-dominated\footnote{Compton dominance is the ratio of the inverse Compton to synchrotron peak luminosities.} SED \citep[][]{4fgl_dr4}. The BLR/torus radiation field, which provides seed photons for the inverse Compton process, can also effectively absorb photons having energy greater than 100 GeV photons through pair production \citep[e.g.,][]{2003APh....18..377D,2016ApJ...821..102B}. These hypotheses explain the paucity of the known VHE-emitting FSRQs in the TeVCat. Only 12 FSRQs have been reported as VHE emitters, including S5~1027+74 which has also been detected at TeV energies \citep[][]{2025ApJ...991L...8P}. Therefore, it is crucial to increase the sample size of VHE-detected broad emission line blazars, study of which can help us constrain the location of the $\gamma$-ray emitting region along the jet, and  also provide vital clues regarding the jet environment where VHE radiation originates.

This work presents the results of a systematic analysis of $\sim$17.5 years of Fermi-LAT data to construct a definitive catalog of VHE-emitting Compton-dominated blazars. The sample details are provided in Section 2. We describe the data reduction methodology in Section 3. In Section 4, we present the results. In Section 5, we discuss individual VHE-emitting FSRQs, and Section 6 summarizes our findings.

\section{Sample}
\label{sec:sample}
\citet[][]{2021ApJS..253...46P} reported a positive correlation between the accretion rate in Eddington units and Compton dominance. Therefore, a large Compton dominance ($>$1) can be considered as a good proxy to select blazars with radiatively efficient accretion, or FSRQs. We did not directly choose sources classified as FSRQs in the Fermi-LAT catalogs, since some FSRQs have been wrongly reported as BL Lacs in them \citep[e.g., S5 1027+74;][]{2025ApJ...991L...8P}. Therefore, we selected 626 sources from the catalog of \citet[][]{2021ApJS..253...46P} whose Compton dominance was reported to be larger than unity. All of them had spectroscopic redshifts, and black hole mass and accretion disk luminosity were reliably measured \citep[][]{2021ApJS..253...46P}. These quantities are crucial to derive the optical depth for the $\gamma\gamma$ pair production due to the BLR photon field and constraining the location of potential VHE radiating emission region \citep[][]{2016ApJ...821..102B,2016ApJ...830...94F}. We analyzed the Fermi-LAT data for all 626 objects adopting the methodology described in the next section.

\section{\textit{Fermi}-LAT Data Reduction}
\label{sec:fermi}

We analyzed Fermi-LAT data in the energy range 100 GeV - 1 TeV observed from 2008 August 5 to 2026 April 7 (MJD 54683-61138). The lower energy threshold of 100 GeV was adopted to focus on the VHE regime, conventionally defined above this energy \citep[e.g.,][]{2022ApJ...933..213D}. The analysis was carried out using the {\tt fermipy} package in conjunction with the standard FermiTools \citep{2017ICRC...35..824W}. Given the low photon counts, we adopted an unbinned likelihood fitting technique for the data analysis.

Events were extracted from a circular region of radius $5^\circ$ centered on the source of interest, selecting SOURCEVETO class photons ({\tt evclass=2048}) that has considerably lower background compared to SOURCE class events\footnote{\url{https://fermi.gsfc.nasa.gov/ssc/data/access/lat/lat\_data\_products.html}}. To ensure the high-quality data, we applied the recommended selection criteria $\tt{DATA\_QUAL > 0}$ and $\tt{LAT\_CONFIG == 1}$. The events were spatially binned with a pixel size of $0.05^\circ$ and divided into 10 logarithmically spaced energy bins per decade. Diffuse background emission was modelled using the Galactic diffuse emission model $\tt{gll\_iem\_v07}$ and the isotropic diffuse model $\tt{iso\_P8R3\_SOURCEVETO\_V3\_v1}$ \citep[][]{2016ApJS..223...26A}. To minimize contamination from Earth limb $\gamma$-ray emission, a zenith angle cut of $z_{\rm max}<105^{\circ}$ was imposed.

The source model was constructed by considering all FL16Y sources lying within a $7^\circ$ region of interest centered on the target. Emission from the known extended \gm-ray sources was accounted for using the extended source templates\footnote{\url{https://fermi.gsfc.nasa.gov/ssc/data/access/lat/fl16y/LAT\_extended\_sources\_16years.tgz}}. Spectral parameters of bright sources with TS $>$ 25 were left free during the unbinned likelihood optimization, while sources with TS $<$ 1 were removed from the model. Owing to the limited photon statistics above 100 GeV, the target of interest was modelled with a simple power-law spectral shape.

The significance of the $\gamma$-ray emission was evaluated using the maximum-likelihood test statistic, defined as $\mathrm{TS} = 2\,\Delta \log$ (L), where L denotes the likelihood values of models with and without the source \citep[][]{1996ApJ...461..396M}. The $\tt{gtsrcprob}$ tool was employed to determine the energy of the highest-energy photon associated with the source, requiring a source-association probability of at least 95\%.

At very high energies, $\gamma$-ray photons are subject to attenuation via pair production interactions with the diffuse Extragalactic Background Light (EBL)\citep{2013APh....43..112D,2015ApJ...812...60B}. To account for this effect, we repeated the spectral fitting using the EBL attenuation model of \citet[][]{2011MNRAS.410.2556D}, and derived the EBL-corrected spectral parameters.

\begin{table*}
	\centering
	\setlength{\tabcolsep}{5.3pt}
	\setlength\extrarowheight{4pt}
        \caption{List of $\gamma$-ray emitting FSRQs detected in VHE band. The upper panel lists sources with Test Statistic (TS) values greater than 25 and the lower panel includes sources with 16$<$TS$<$25. The column details are as follows: (1) 4FGL name (counterpart name); (2) redshift; (3) Compton dominance; (4) observed $\gamma$-ray photon flux (in $10^{-12}\,\mathrm{ph}\,\mathrm{cm}^{-2}\,\mathrm{s}^{-1}$); (5) photon index; (6) TS; (7),(8) and (9) EBL corrected $\gamma$-ray photon flux(in $10^{-12}\,\mathrm{ph}\,\mathrm{cm}^{-2}\,\mathrm{s}^{-1}$), photon index and TS; (10) whether the source is present in TeVCat}
	\label{tab:source details}
	\begin{tabular}{lccccccccc}
		\hline 
		4FGL Name & z & CD & $Flux_{obs}$ & $\Gamma$$_{obs}$ & $TS_{obs}$ & $Flux_{EBL}$ & $\Gamma_{EBL}$ & $TS_{EBL}$ & TeV \\
		~[1] & [2] & [3] & [4] & [5] & [6] & [7] & [8] & [9] & [10] \\
		\hline
        J0102.8+5824 (TXS 0059+581) & 0.644 & 1.48&$5.2\pm2.3$&$5.2\pm1.9$&     42&$5.2\pm2.3$&  $3.4\pm2.1$&43& N\\
        J0348.6-2749 (PKS 0346-27) &0.991&3.31&$0.7\pm0.6$&$17.8\pm8.6$&41&   --&            -- &           --  &Y\\
        J0428.6-3756 (PKS 0426-380)  &1.110&6.31&$3.9\pm2.3$&$6.9\pm3.4$&35&$3.9\pm2.3$&$3.9\pm3.8$&35&N\\
        
        J0538.8-4405 (PKS 0537-44) &0.896&1.58&$4.9\pm2.5$&$5.7\pm2.4$&37&  $4.9\pm2.5$&$3.2\pm2.7$&38&N\\
        
        J0904.9-5734 (PKS 0903-57) &0.262&2.88&$11.6\pm3.8$&$2.9\pm0.8$&82&  $11.6\pm3.8$&$2.1\pm0.8$&81&Y\\
        
        J0957.6+5523 (4C +55.17) &0.903&8.13&$3.9\pm2.1$&$4.8\pm2.1$&34&  $3.9\pm2.1$&$1.9\pm2.4$&35&N\\
        
        J1031.1+7442 (S5 1027+74) &0.123&1.10&$5.8\pm2.5$&$2.2\pm0.8$&39&  $5.8\pm2.5$&$1.8\pm0.8$&40&N\\
        
        J1224.9+2122 (4C +21.35) &0.434&5.62&$5.9\pm2.9$&$3.1\pm1.2$&31&  $5.9\pm2.9$&$1.6\pm1.4$&31&Y\\
         
        J1310.5+3221 (OP 313) &0.996&2.45&$4.6\pm2.9$&$8.9\pm4.5$&    62&$6.5\pm2.9$&$9.7\pm4.9$&69&Y\\
        
        J1512.8-0906 (PKS 1510-089) &0.362&13.49&$12.6\pm4.6$&$3.2\pm0.9$&     55&$12.4\pm4.6$&$2.0\pm1.1$&56&Y\\
           
		\hline
        J0457.0-2324 (PKS 0454-234) &1.003&10.00&$2.8\pm1.9$&$15.8\pm3.7$&17&  $2.8\pm1.9$&$13.9\pm10.6$&17&N\\
        J1058.4+0133 (4C +01.28) &0.891&1.66&$5.4\pm3.2$&$5.0\pm2.2$&19&$5.5\pm3.2$&$2.2\pm2.6$&20&N\\        
        J1159.5+2914 (Ton 599) &0.725&2.34&$3.2\pm2.3$&$5.5\pm3.4$&16&  $3.3\pm2.3$&$3.4\pm3.9$&17&Y\\

        J1256.1-0547 (3C 279) &0.536&1.95&$5.7\pm3.2$&$5.2\pm3.0$&23&  $5.6\pm3.3$&$3.9\pm3.8$&23&Y\\
        
        \hline
	\end{tabular}
\end{table*}

\begin{table*}
	\centering
	\setlength{\tabcolsep}{25pt}
    \renewcommand{\arraystretch}{1.2}
        \caption{List of FSRQs with at least one photon detected above 100 GeV with $\ge$95$\%$ source probability. The upper panel lists sources with TS $>$ 25, middle panel includes sources with TS values in the range 16$<$TS$<$25 and lower panel shows sources with TS$<$16. The column details are as follows: (1) 4FGL name; (2) redshift; (3) Compton dominance; (4) number of VHE photons detected; (5) Energy of highest energy photon (in GeV); (6) Time of highest VHE photon detection (in MJD)}.
	\label{tab:all sources}
	\begin{tabular}{lccccc}
		\hline 
		4FGL Name & z & CD & No. & $E_{HEP}$ & $T_{arrival}$  \\
		\hline
        J0102.8+5824 & 0.644&1.48 & 5 & 152.91&57190.78\\
        J0348.6-2749&0.991&3.31&3&103.55&58550.54\\
        J0428.6-3756&1.110&6.31&3 &122.05&55209.12\\
        J538.8-4405&0.896&1.58&4&145.72&57895.44\\
        J0904.9-5734&0.262&2.88&9&480.78&59553.44\\
        J0957.6+5523&0.903&8.13&3&145.88&55115.65\\
        J1031.1+7442&0.123&1.10&5&341.77&57193.29\\
        J1224.9+2122&0.434&5.62&4&256.55&59718.29\\
        J1310.5+3221&0.996&2.45&5&128.48&60272.73\\
        J1512.8-0906&0.362&13.49&6&230.84&56056.97\\
        \hline
        J0457.0-2324&1.003&10.00&2&111.02&55222.53\\        
        J1058.4+0133& 0.891&1.66& 2& 120.25& 60879.43\\
        J1159.5+2914&0.725&2.34&2&139.85&59573.62\\        
        J1256.1-0547&0.536&1.95&3&128.47&56791.54\\
        \hline
        J0043.8+3425&0.969&12.30&3&158.78&57468.78\\
        J0221.1+3556&0.954&5.13&1&101.52&56851.78\\
        J0405.6-1308&0.571&2.45&1&102.04&57396.89\\
        J0433.6+2905&0.910&12.59&1&110.14&60177.08\\
        J0608.0-0835&0.873&2.51&1&175.27&56707.11\\
        J0739.2+0137&0.191&1.70&1&105.62&57072.61\\
        J0803.2-0337&0.365&2.14&2&109.86&54698.99\\
        
        J0811.4+0146& 1.148&2.19& 1 &113.83& 55505.87\\
        J1127.0-1857&1.052&1.55&1&114.56&56327.83\\        
        J1322.2+0842&0.326&1.05&1&126.31&54758.03\\
        J1337.6-1257&0.545&1.15&1&139.79&55027.12\\
        J1422.4+3223&0.682&1.41&1&146.36&58865.67\\
        J1647.5+4950& 0.047&1.12& 1& 131.05& 59594.40\\
        J1503.5+4759& 0.345&2.69& 1 &252.78& 56029.13\\
        J1700.0+6830&0.301&1.35&1&135.78&60738.28\\
        J1722.7+1014&0.732&1.07&1&171.86&54972.28\\
        J2000.9-1748& 0.654&1.70 &1& 110.67& 55340.01\\
        J2143.5+1743&0.213&4.47&1&182.53&56700.86\\
        J2211.2-1325&0.392&2.24&1&137.39&55484.74\\
        J2253.9+1609&0.859&10.23&1&143.39&60092.59\\
        J2326.2+0113&1.600&1.32&1&177.78&55307.04\\
        \hline

	\end{tabular}
\end{table*}

\begin{figure*}
\hbox{
    \includegraphics[scale=0.3]{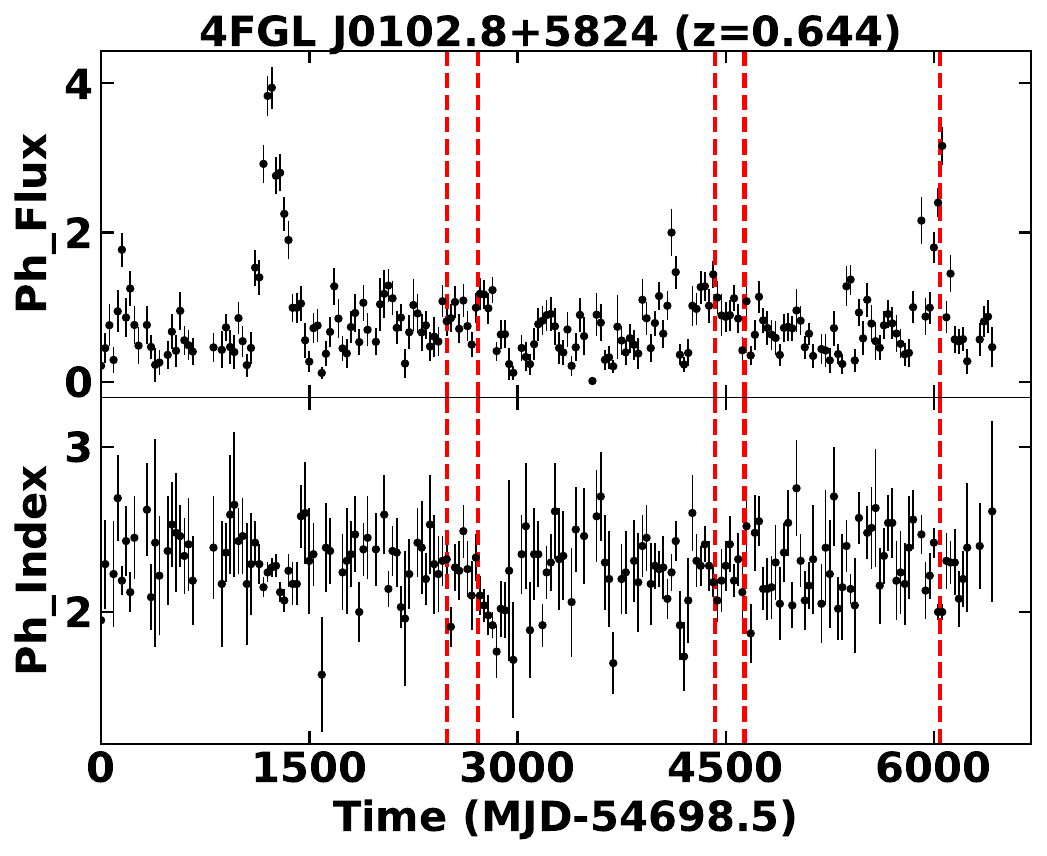}
    \includegraphics[scale=0.3]{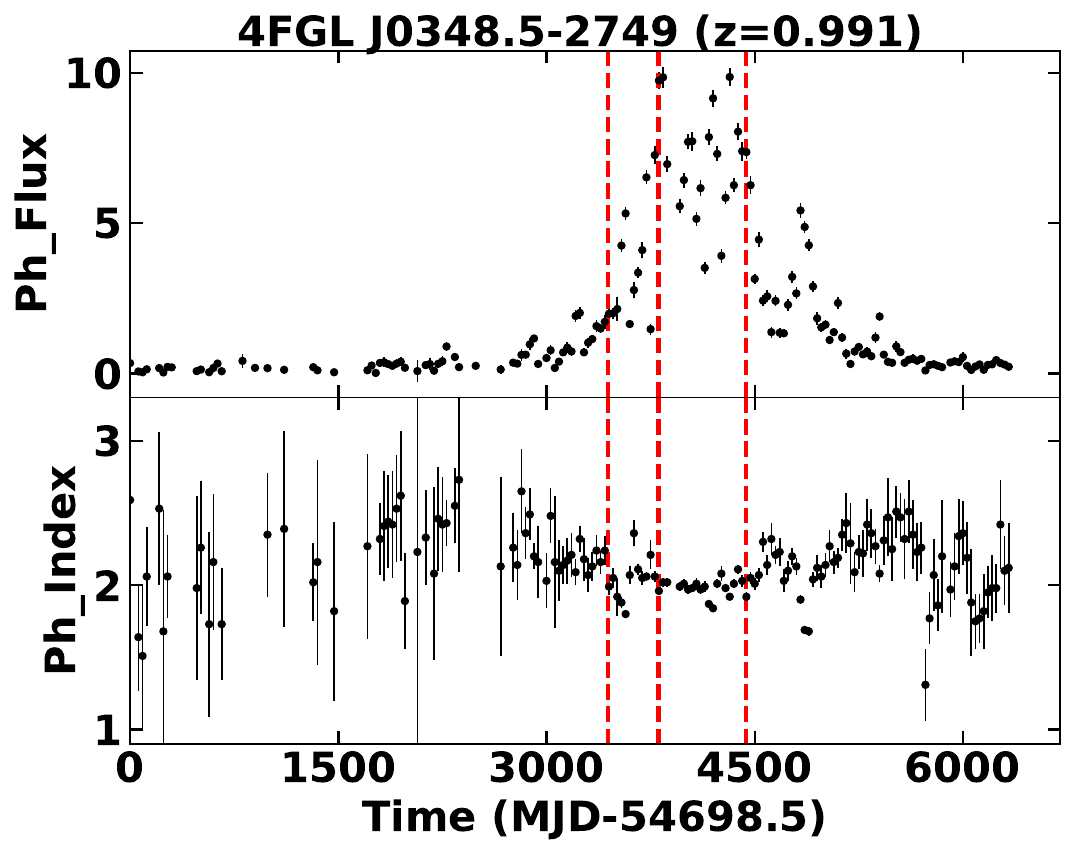}
    \includegraphics[scale=0.3]{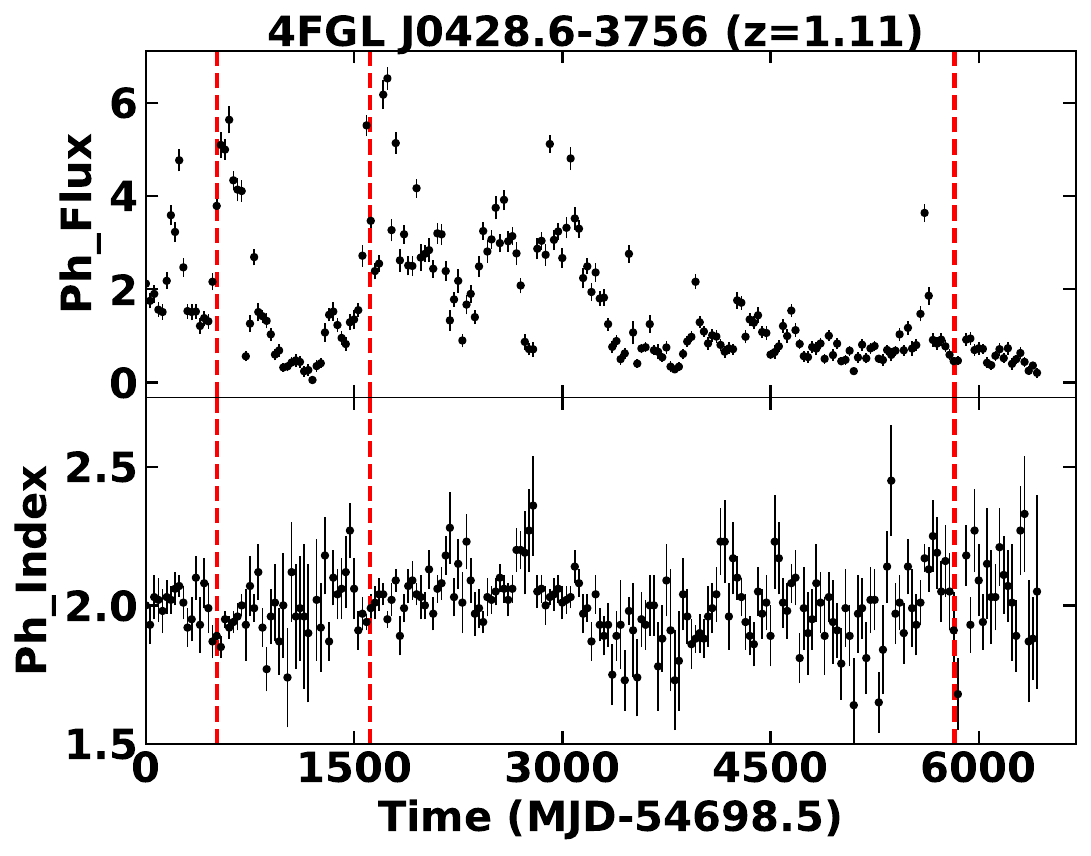}
     }
\hbox{
    \includegraphics[scale=0.3]{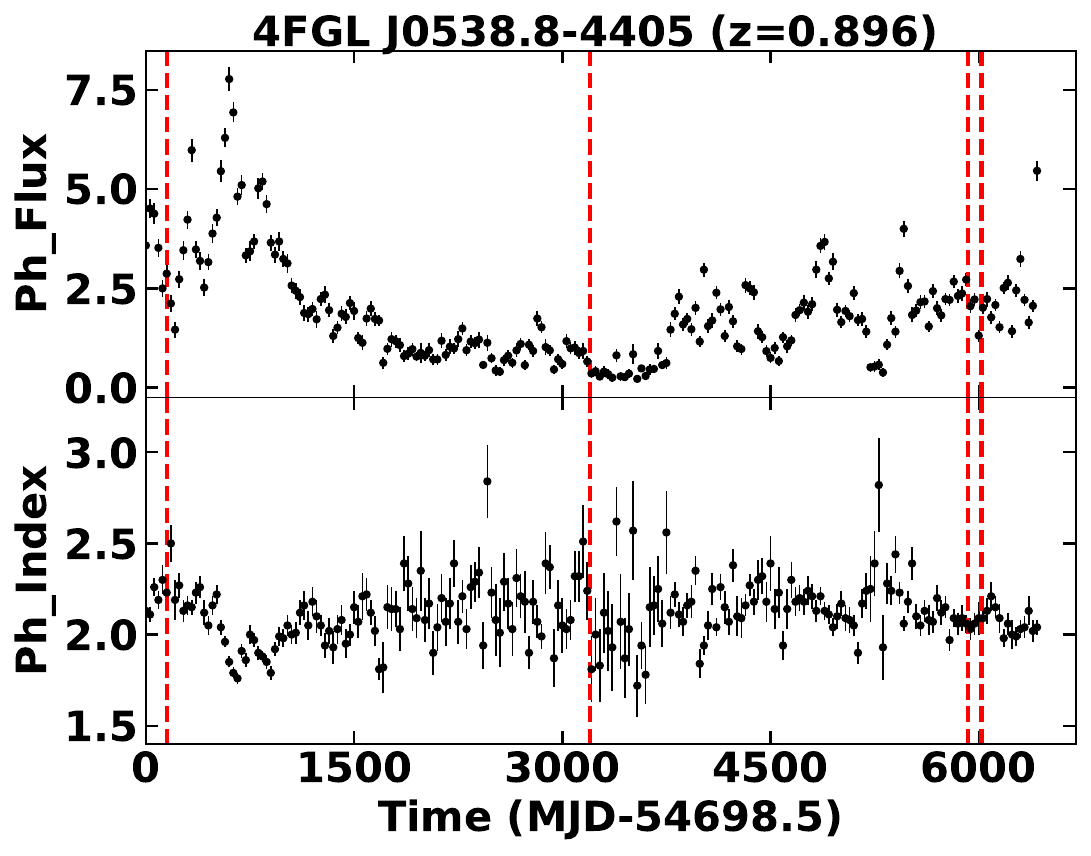}
    \includegraphics[scale=0.3]{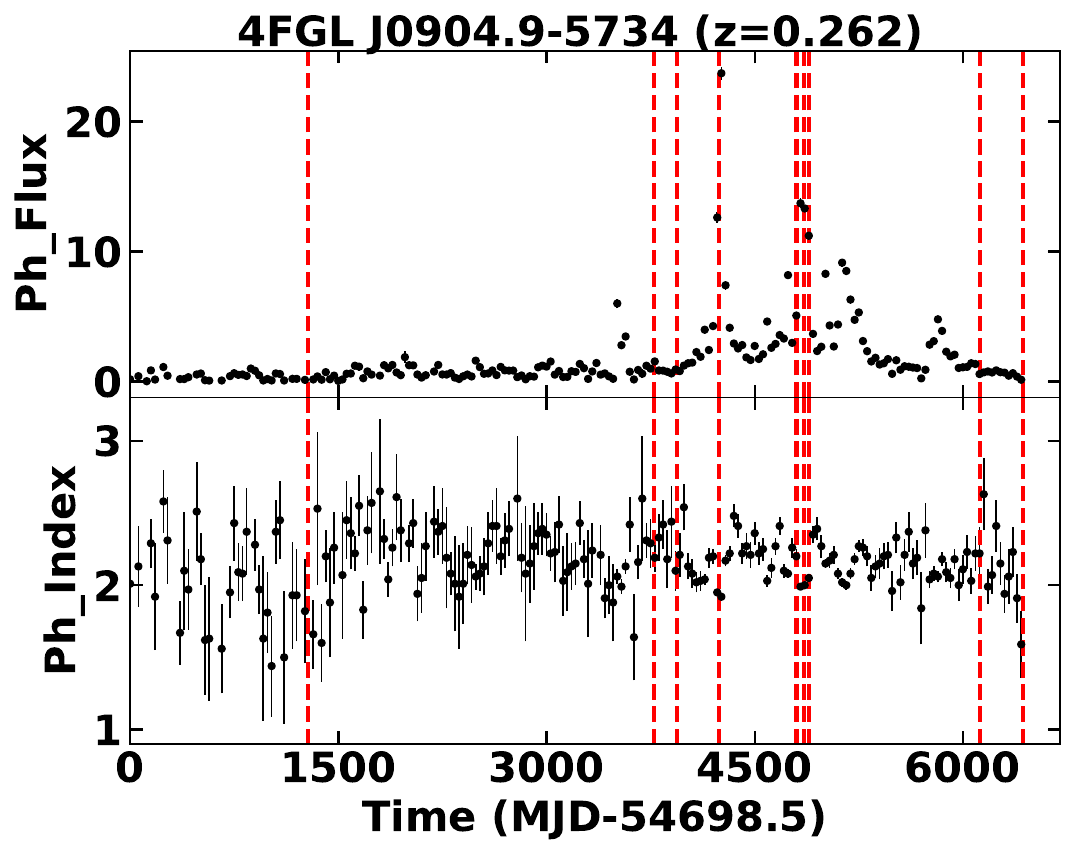}
    \includegraphics[scale=0.3]{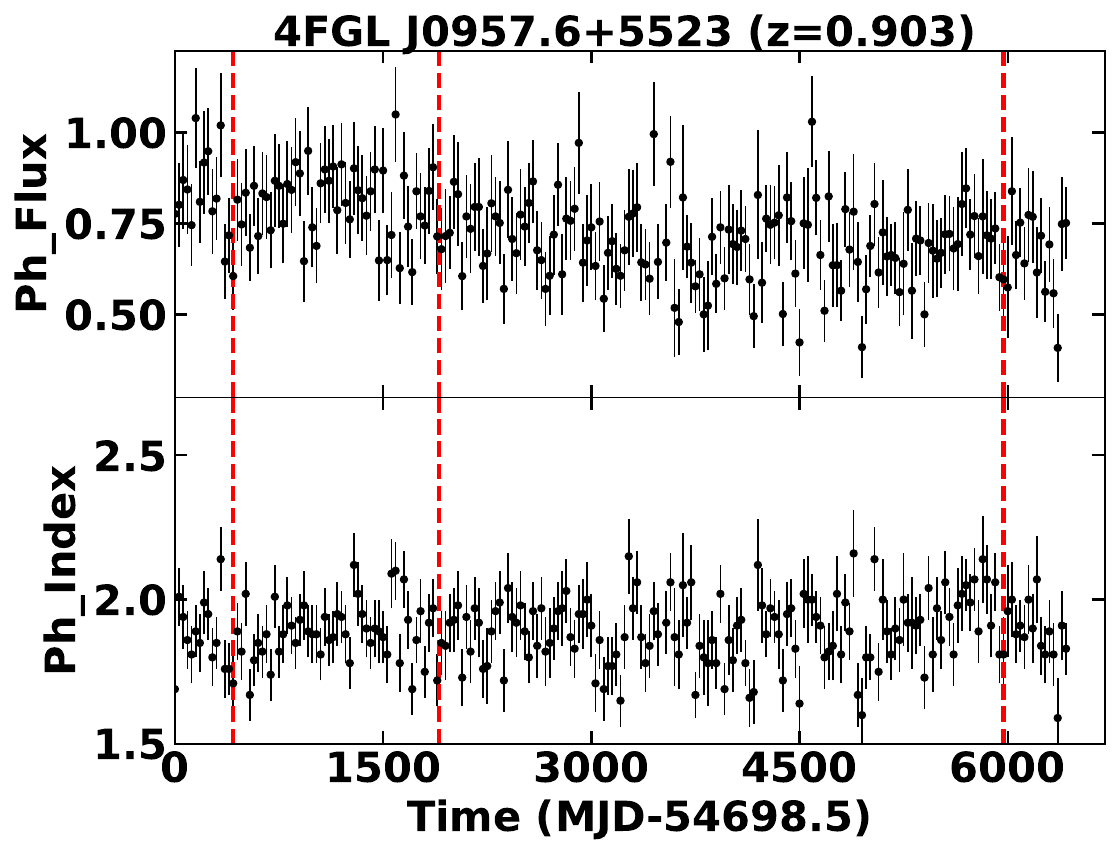}
     }
\hbox{
    \includegraphics[scale=0.3]{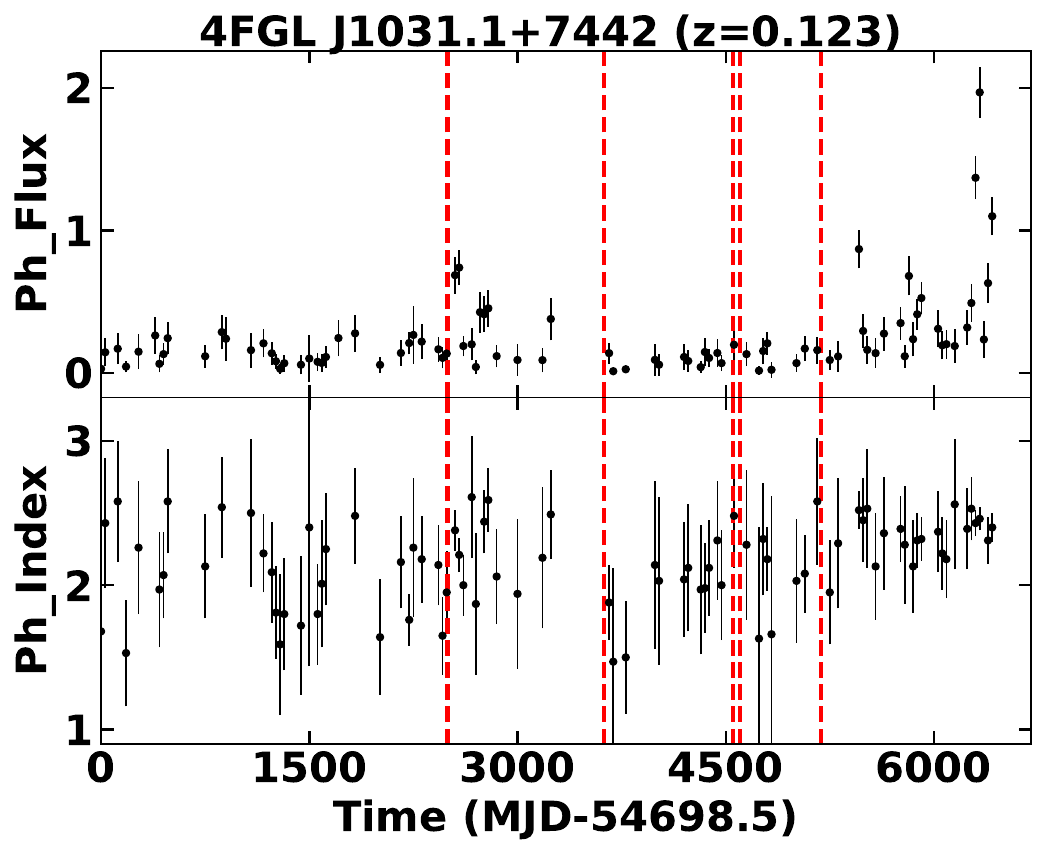}
    \includegraphics[scale=0.3]{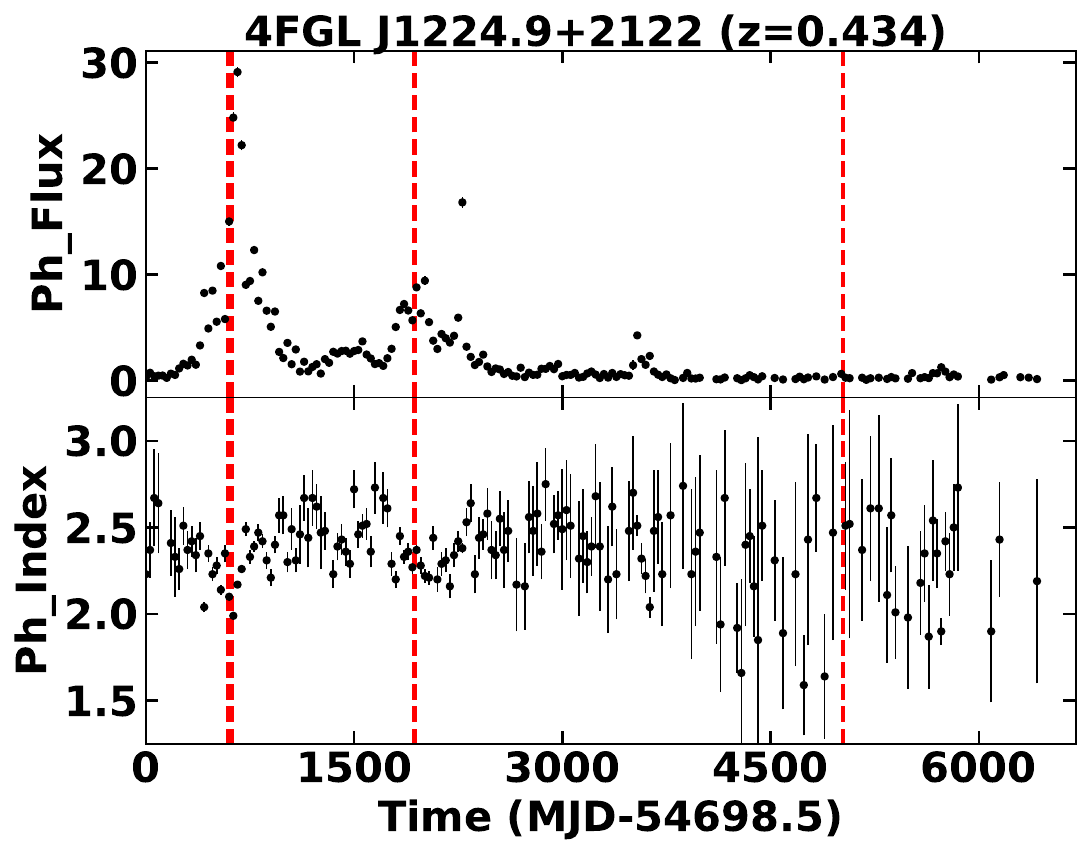}
    \includegraphics[scale=0.3]{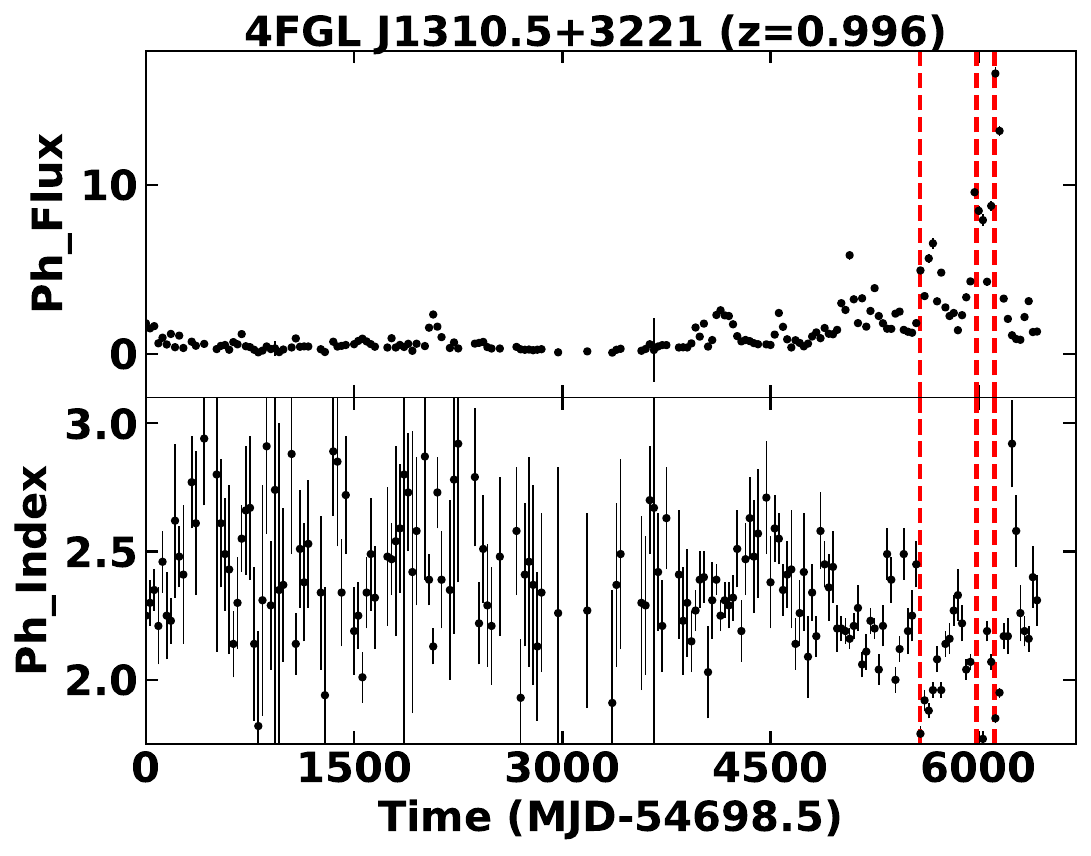}
     }
\hbox{\hspace{5.5cm}
    \includegraphics[scale=0.3]{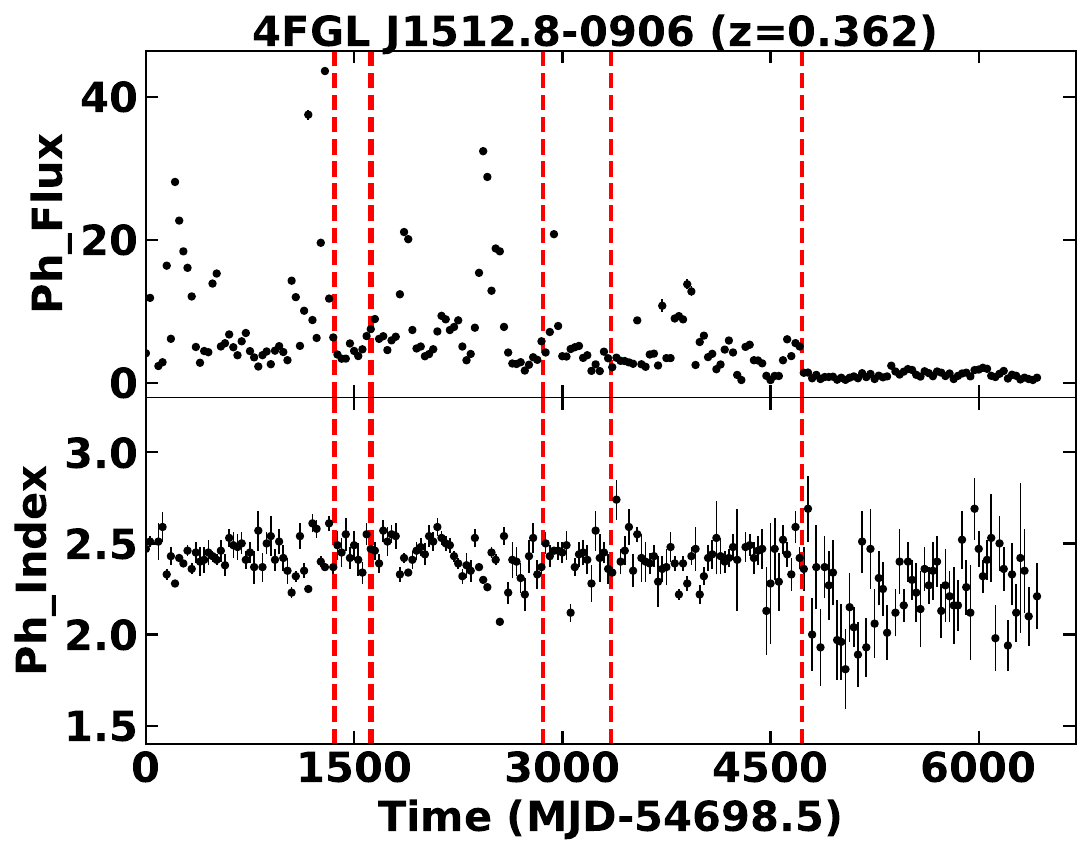}
     }
\caption{Monthly-binned light curves of VHE sources (TS$>$25) in the energy range 0.1$-$100 GeV constructed using the data taken from Fermi-LAT Light Curve Repository. In each plot, top panel represents the photon flux in units of ${10^{-7}}$ ph ${\ cm^{-2}\ s^{-1}}$, and bottom panel refers to the photon index. The red dashed vertical lines represent the time of arrival of VHE photons.}\label{lightcurves}
\end{figure*}

\begin{figure*}
\hbox{
    \includegraphics[scale=0.28]{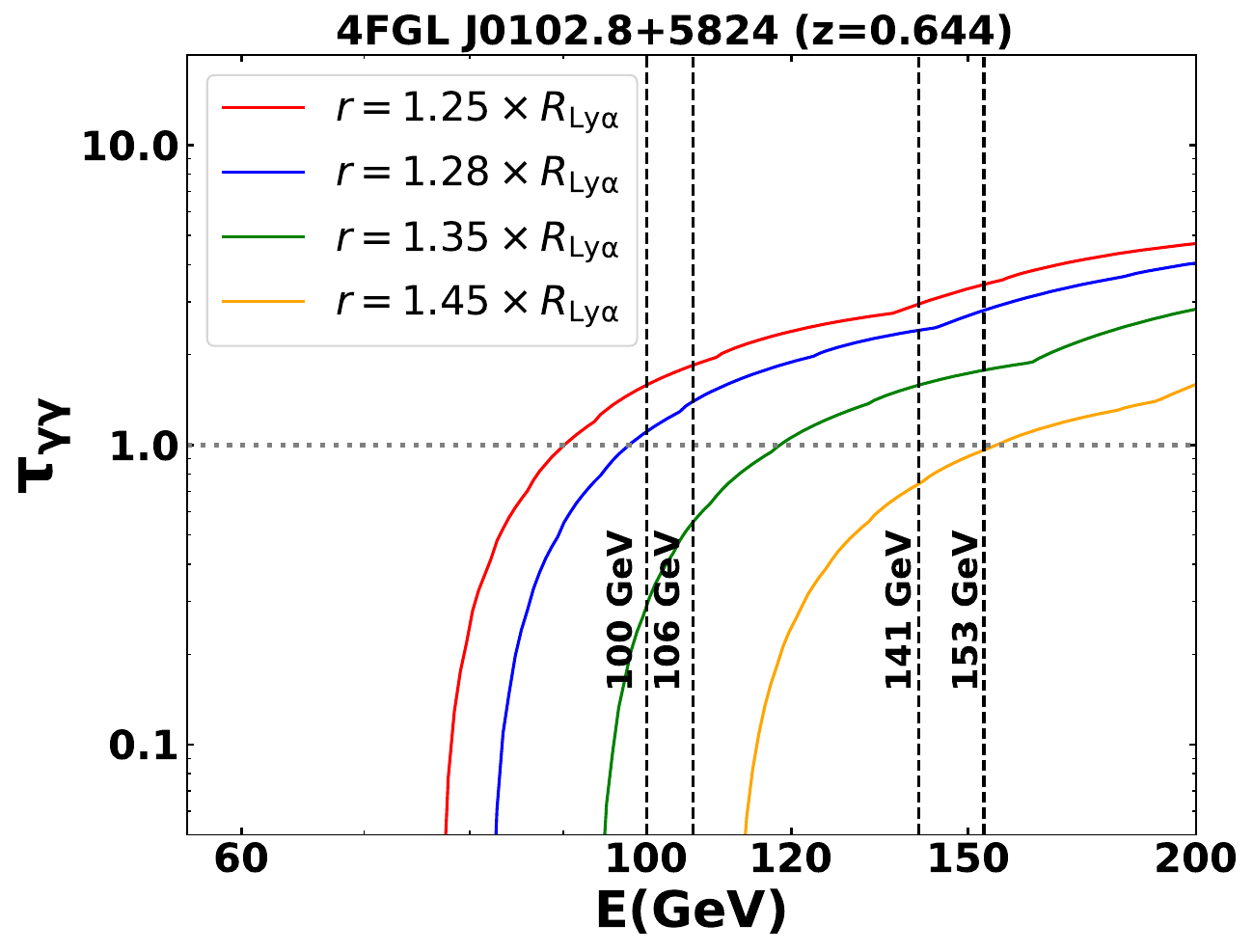}
    \includegraphics[scale=0.28]{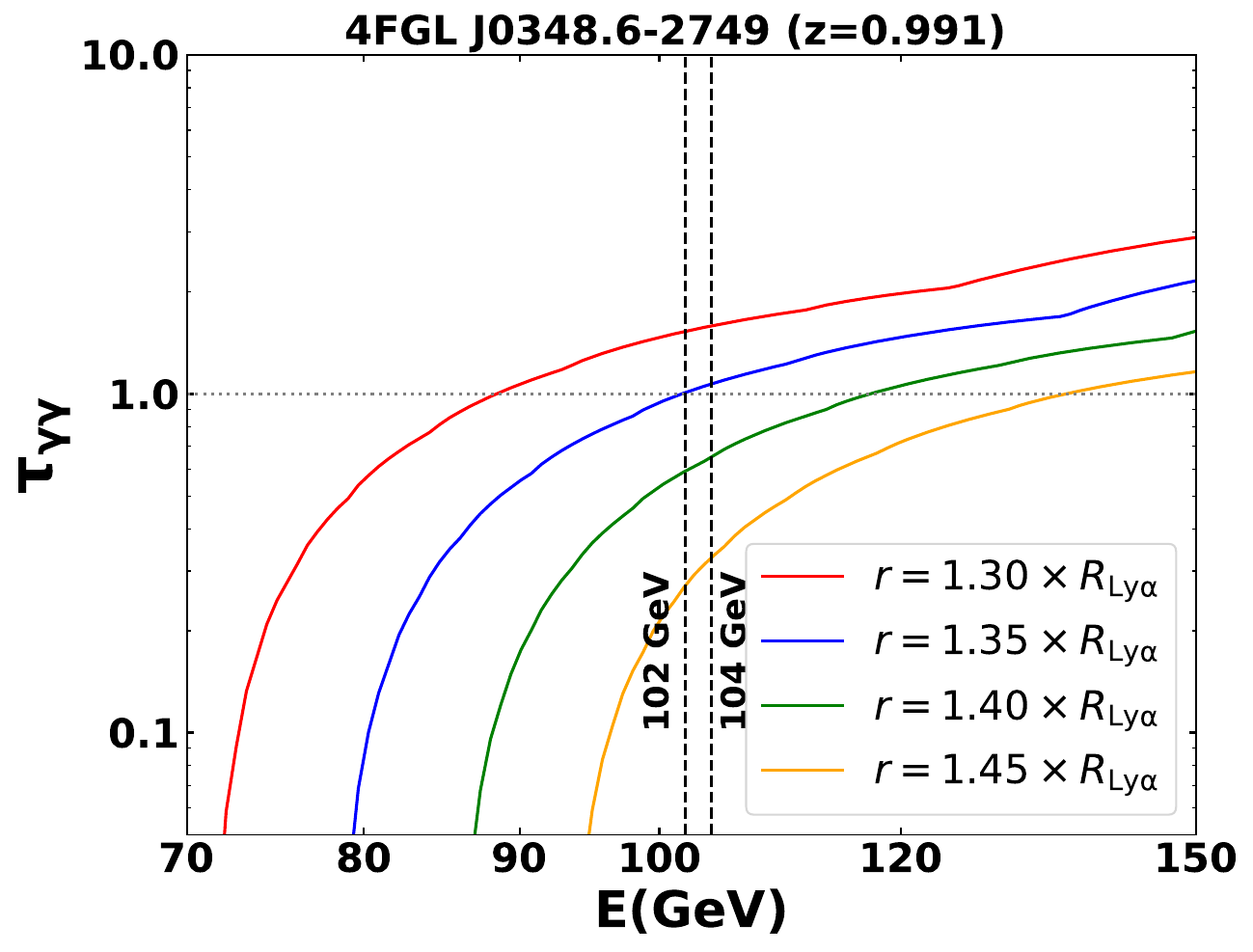}
    \includegraphics[scale=0.28]{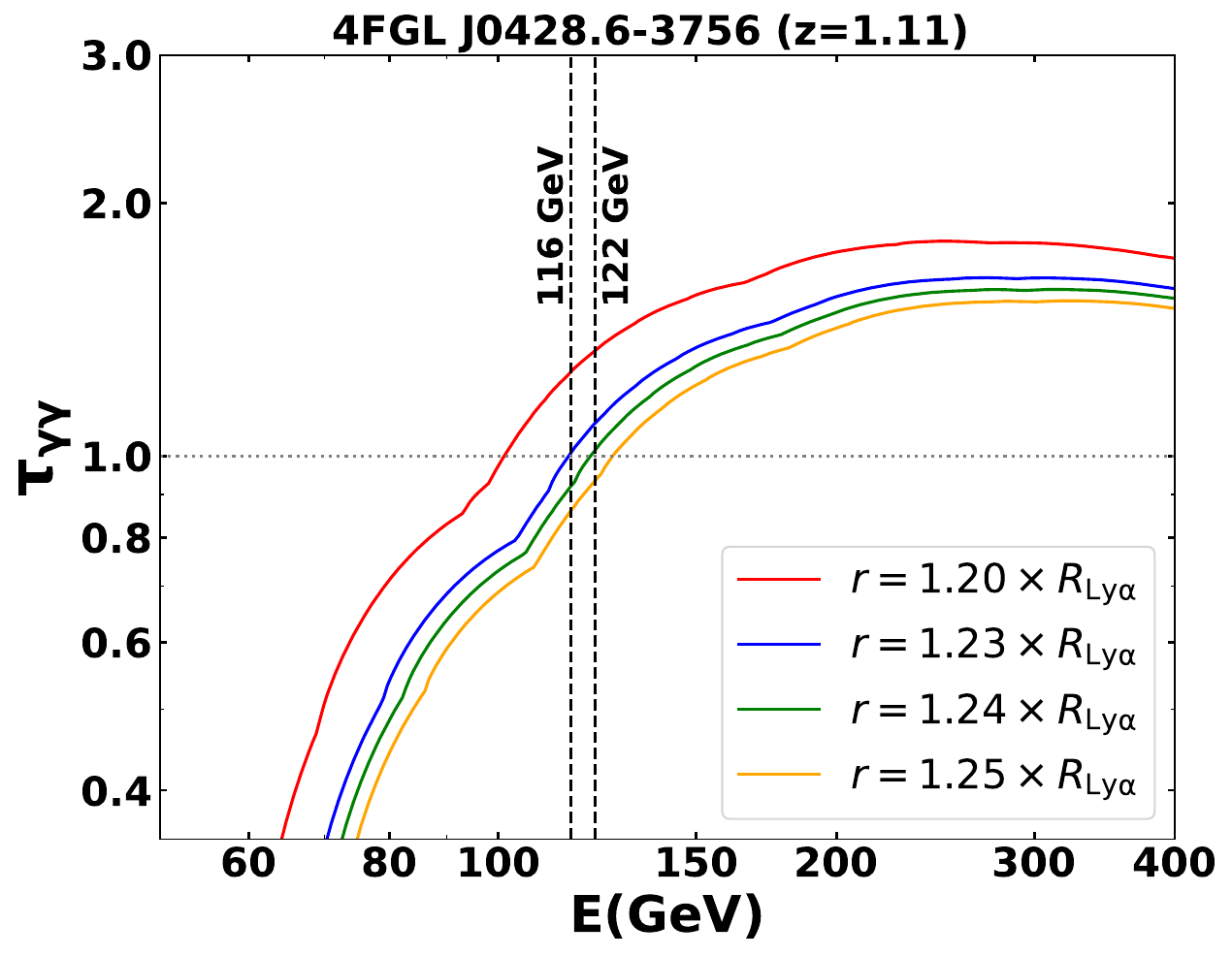}
     }
\hbox{
    \includegraphics[scale=0.29]{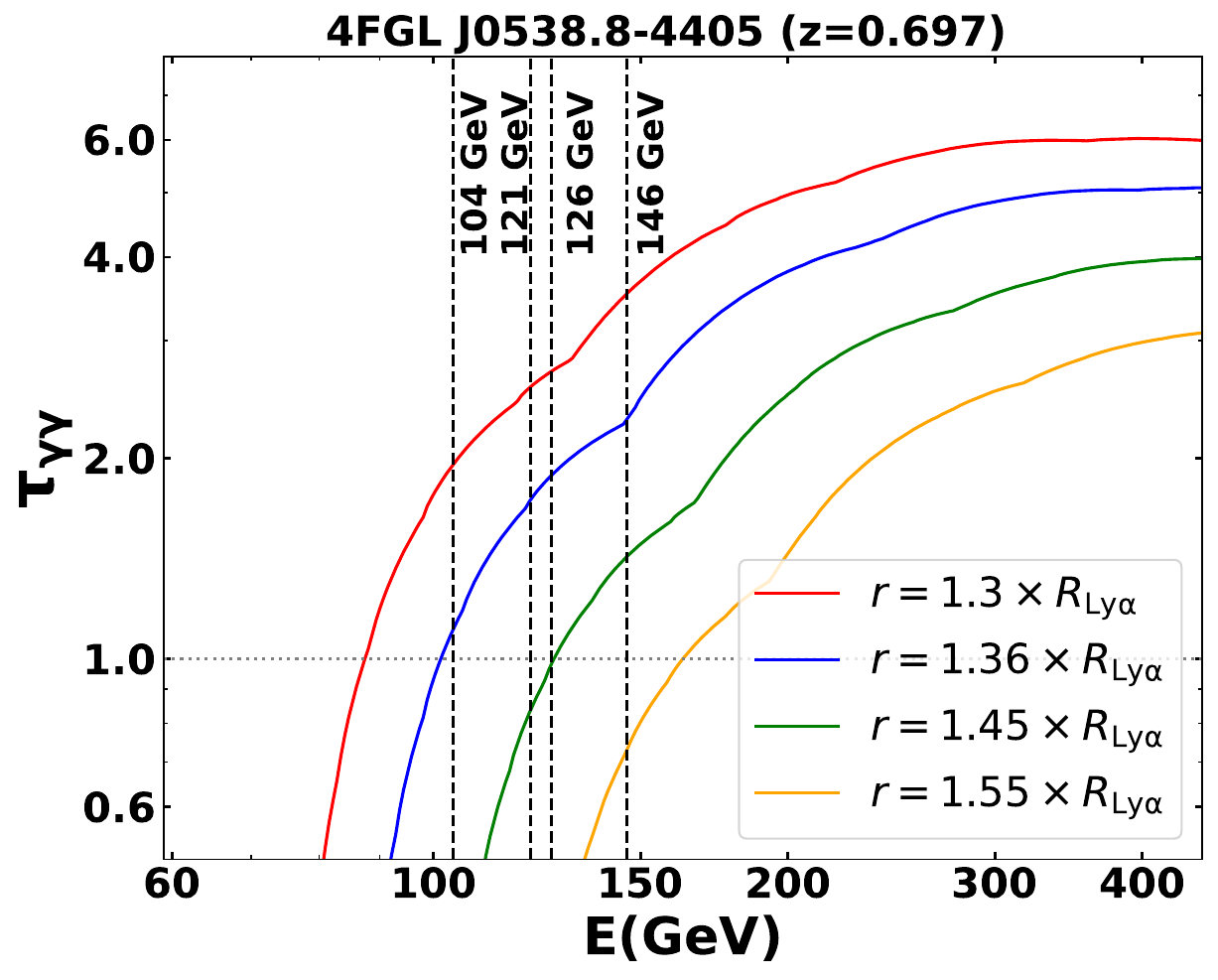}
    \includegraphics[scale=0.29]{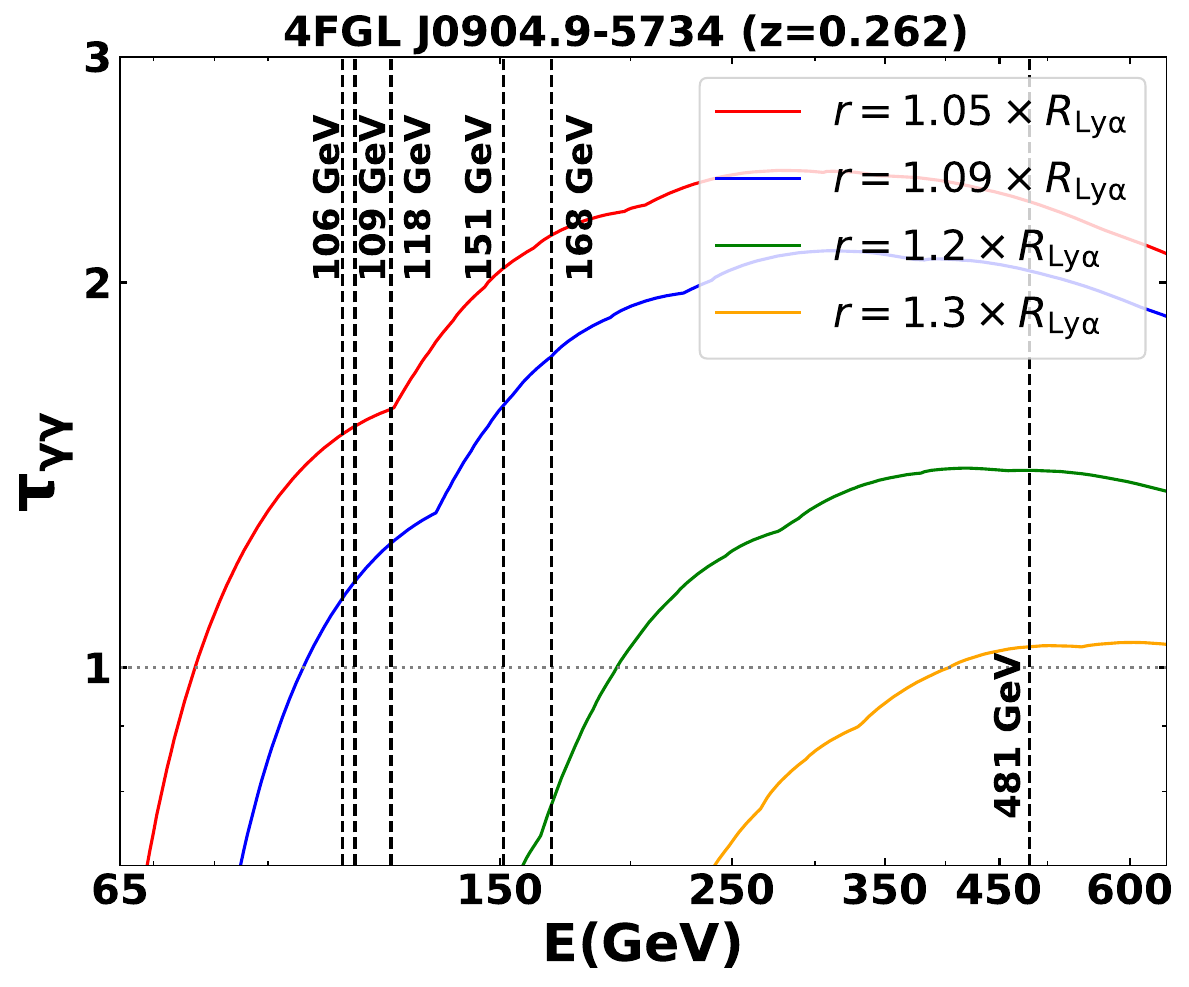}
    \includegraphics[scale=0.29]{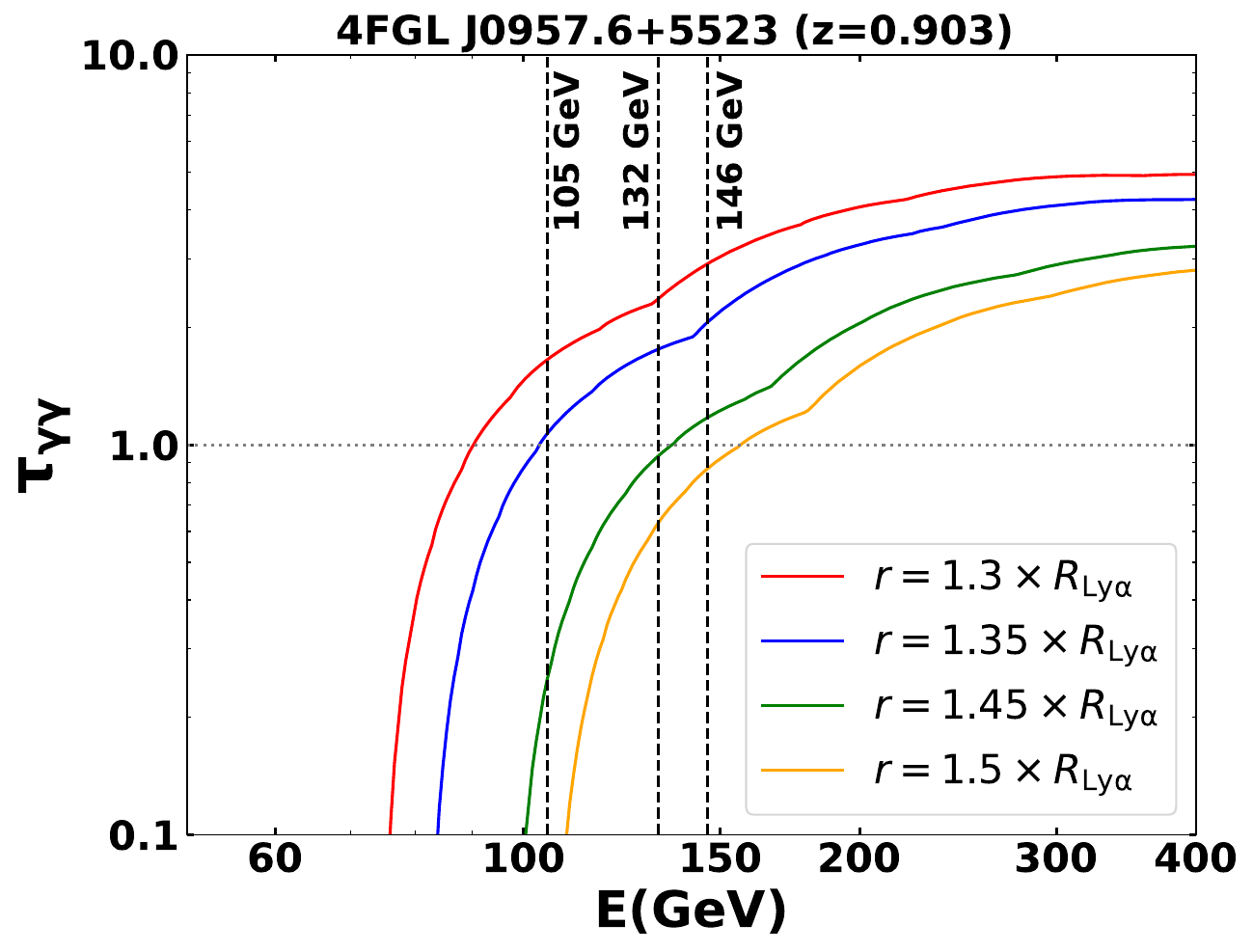}
     }
\hbox{
    \includegraphics[scale=0.28]{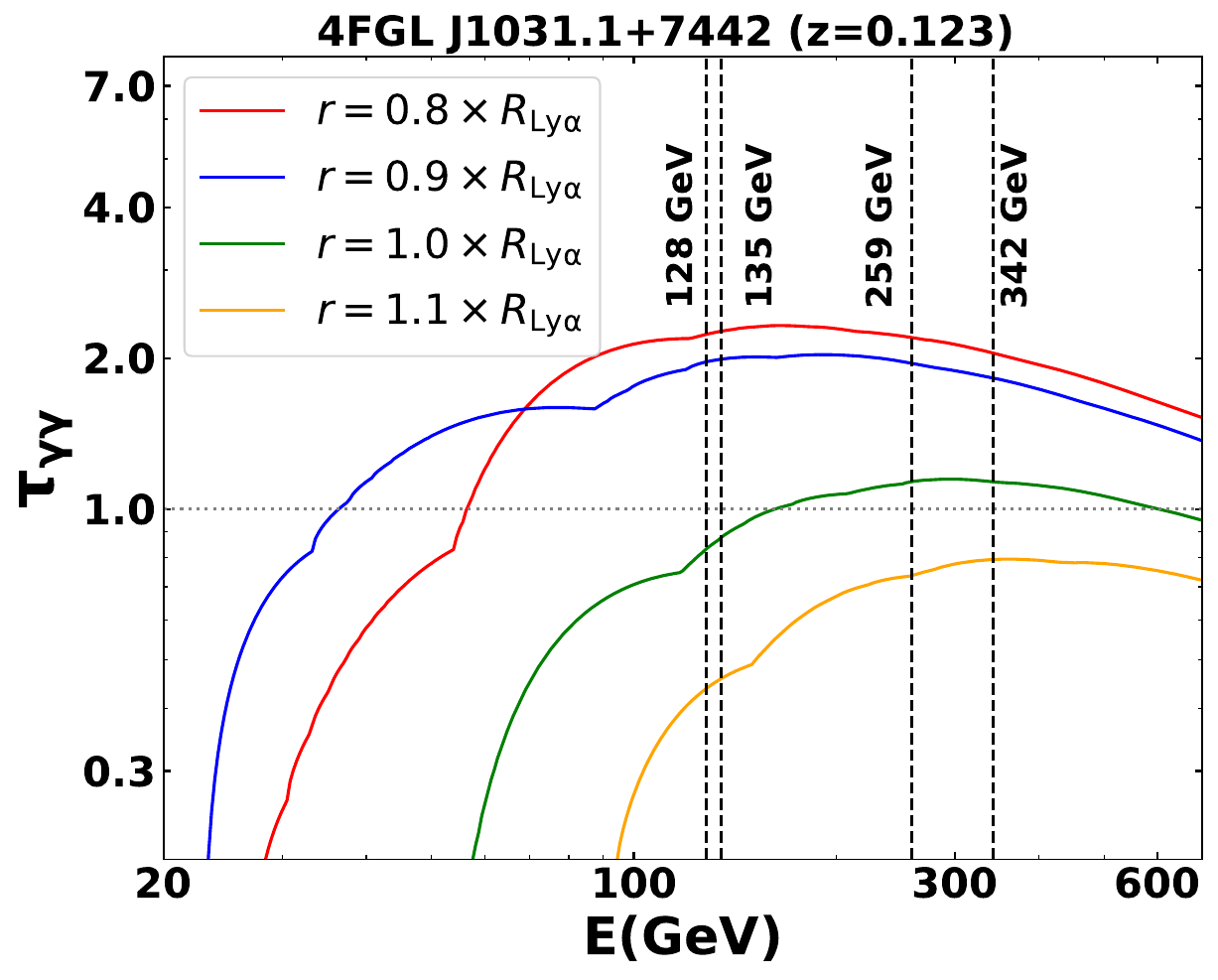}
    \includegraphics[scale=0.28]{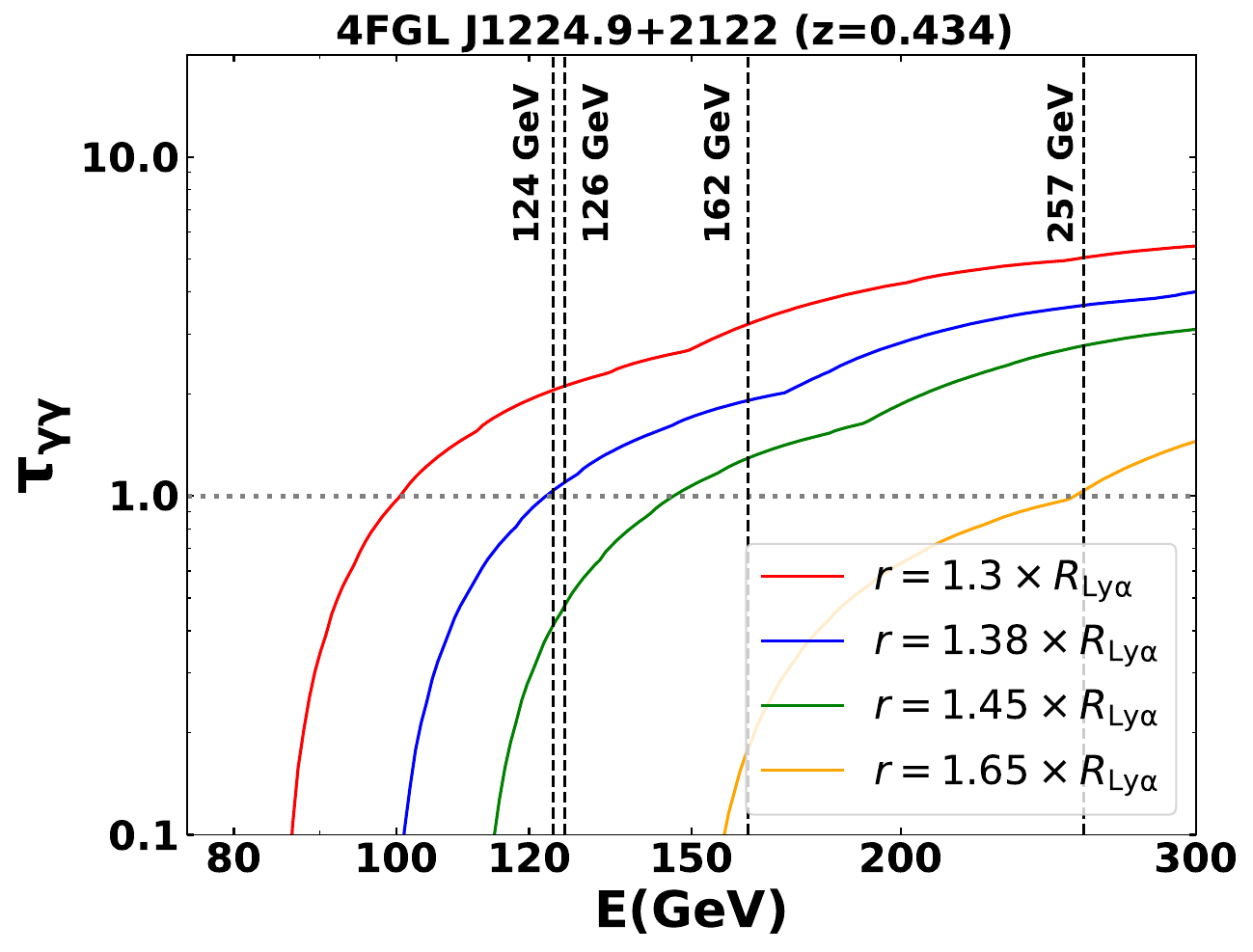}
    \includegraphics[scale=0.28]{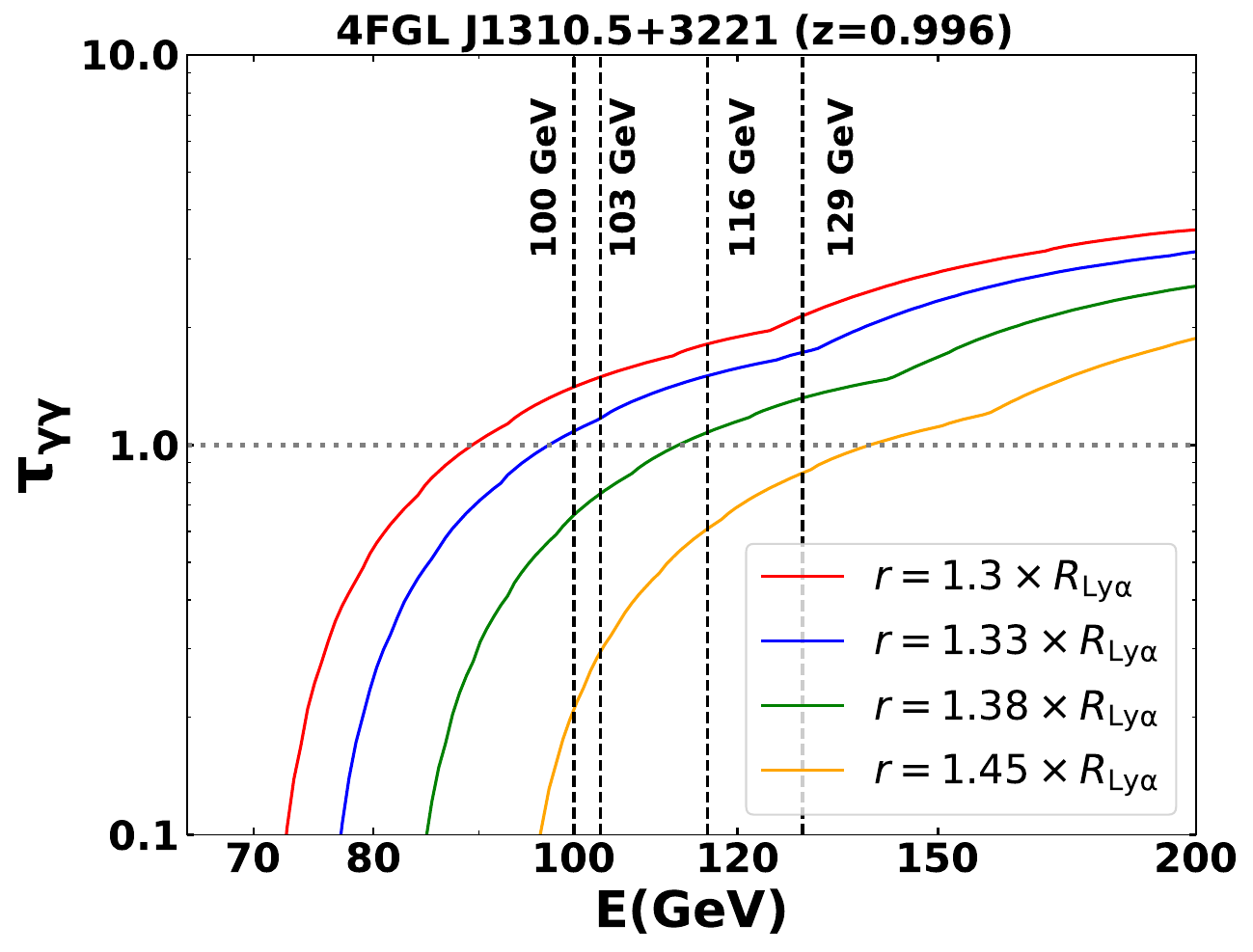}
     }
\hbox{\hspace{5.5cm}
    \includegraphics[scale=0.28]{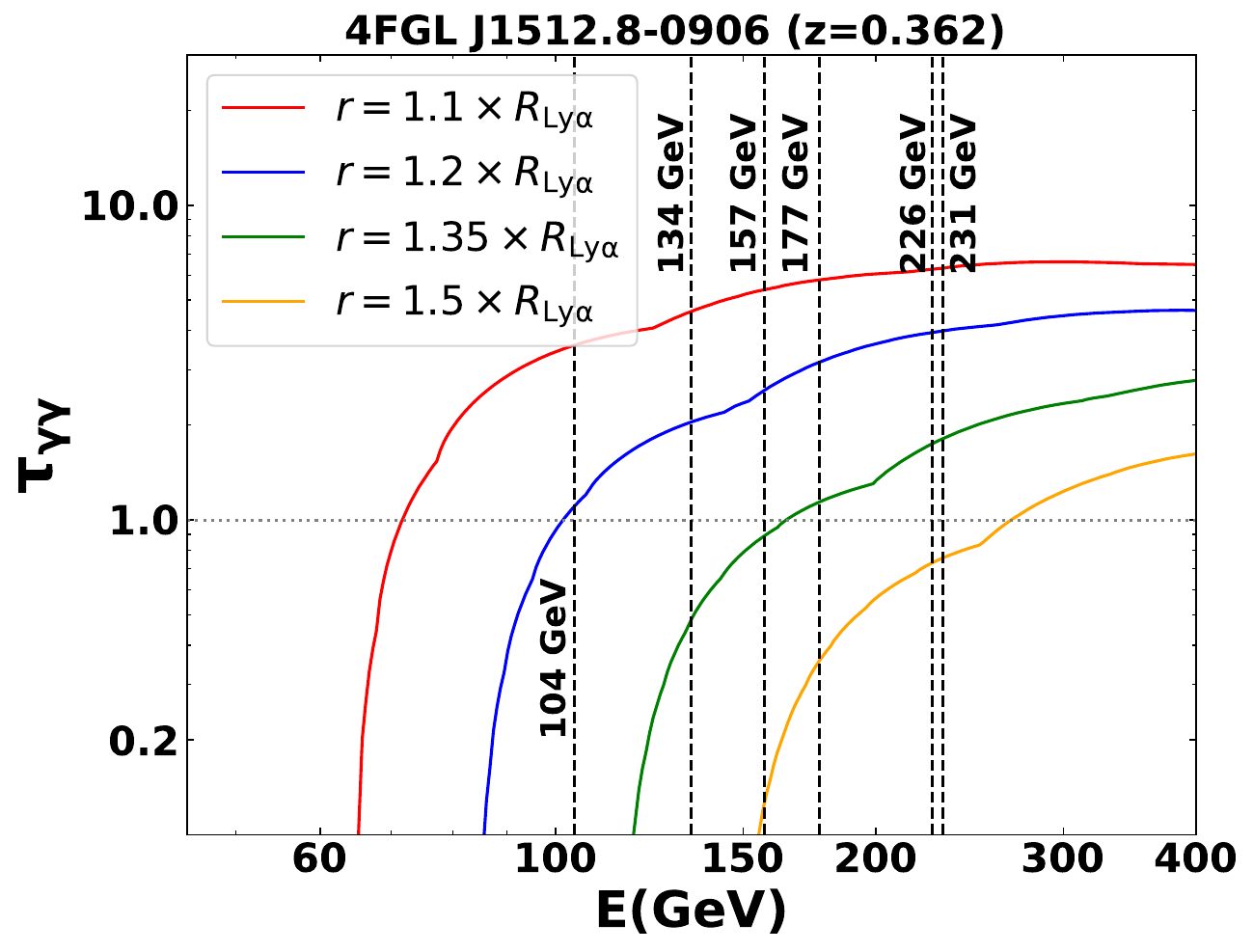}
     }
\caption{Optical depth due to $\gamma\gamma$ pair production with the BLR photons as a function of $\gamma$-ray photon energy for different distances of the emission region from the photon field. Energies of detected VHE photons are highlighted by the vertical dotted lines. See the text for details.}\label{opacity}
\end{figure*}

\begin{figure}
\centering
\includegraphics[width=0.9\columnwidth]{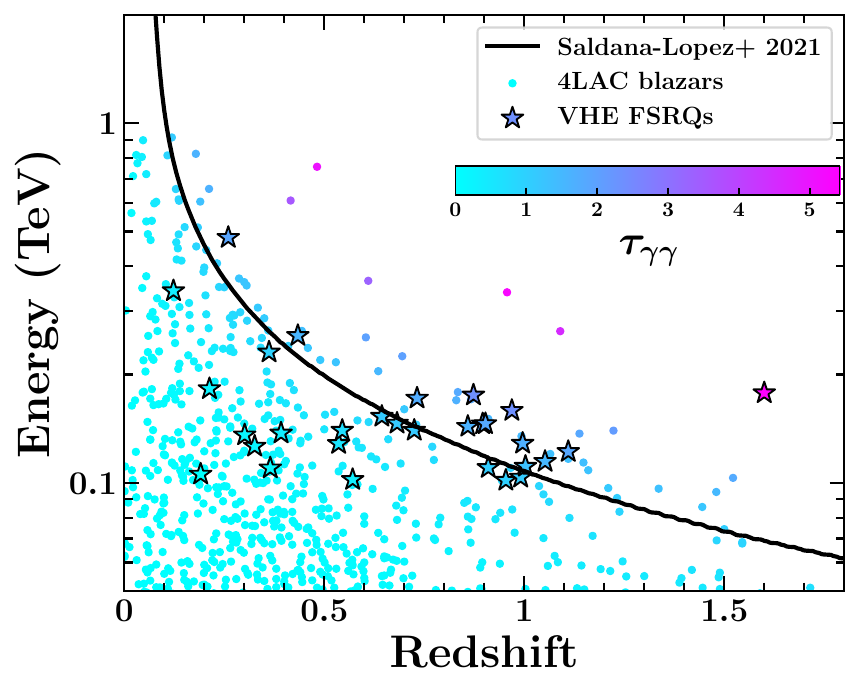}
		
\caption{Cosmic $\gamma$-ray horizon. The black solid line indicates the opacity regime corresponding to an EBL optical depth of $\tau_{EBL} = 1$ \citep{2021MNRAS.507.5144S}. Source markers are color coded by their respective EBL optical depth values derived from the same model. The VHE-emitting FSRQs are shown as stars, whereas, filled circles refer to blazars present in the fourth catalog of the Fermi-LAT detected AGN \citep[4LAC;][]{2020ApJ...892..105A}.}
	\label{Cosmic horizon}	
\end{figure}

\section{Results and Discussion}
\label{sec:results}
\subsection{VHE Detection}
The Fermi-LAT data analysis of 626 Compton-dominated blazars led to the identification of 14 sources with TS $>16$. Of these detections, ten had TS values greater than 25, i.e., at confidence level greater than $\approx$
$5\sigma$. Of these 14 sources, seven had also been detected from observations using ground-based Cherenkov telescopes, and one source, 4FGL~J1031.1+7442, was reported as a VHE emitter by \citet{2025ApJ...991L...8P}. Therefore, in this work, we report the first VHE detection of 4 sources with high confidence (TS $>25$) and tentative VHE detection of 2 Compton-dominated blazars. All TS $>25$ sources have at least 3 VHE photons ($>95\%$ association probability) detected, thus robustly confirming them to be VHE emitters \citep[see, e.g.,][]{2013ApJ...777L..18T}. Among the significantly detected blazars, \mbox{$\mathrm{4FGL\ J0428.6{-}3756}$} is the most distant FSRQ at $z=1.11$. The derived spectral parameters, observed and EBL-corrected, for all blazars are provided in Table~\ref{tab:source details}. 

We found another 21 sources with at least one VHE photon with source association probability greater than 95\%(Table~\ref{tab:all sources}), though their overall detection significance remained low (TS$<16$). Considering a higher association probability threshold of 99\%, we have 14 objects. We caution that, given the large parent sample size, it is possible that a few of them maybe spurious detections. However, three objects, 4FGL J0221.1+3556 (S3 0218+35), 4FGL J0739.2+0137 (PKS 0736+017), and 4FGL J1422.4+3223 (B2 1420+32), have also been detected by the ground-based Cherenkov telescopes \citep[][]{2016A&A...595A..98A,2020A&A...633A.162H,2021A&A...647A.163M}. Two known VHE-detected FSRQs, 4FGL J1229.1+0202 (3C 273) and 4FGL J1443.9+2501 (PKS 1441+25) are missing in our list. The blazar 3C 273 was recently detected in the VHE band by integrating $\sim$17 years of the VERITAS data\footnote{\url{https://indico.cern.ch/event/1258933/contributions/6491204/attachments/3104111/5500883/Benbow_ICRC2025.pdf}}. The VHE flux above 350 GeV was reported to be $(2.3 \pm 0.5) \times 10^{-13}$ ph cm$^{-2}$ s$^{-2}$, which is fainter than the faintest source detected in our analysis. Moreover, no photon with energy above 100 GeV was detected from this object by the Fermi-LAT. The other source, PKS 1441+25, was detected in the VHE band during its 2015 $\gamma$-ray flare \citep[][]{2015ApJ...815L..22A,2015ApJ...815L..23A}. However, no photon of energy greater than 100 GeV was detected with the Fermi-LAT during this period \citep[][]{2015ApJ...815L..23A}. Based on our Fermi-LAT data analysis also, no VHE photon was detected from this object, thus explaining its absence from the list of VHE-detected blazars reported in this work.

\subsection{VHE Emission and Source Variability}

VHE observations of jetted AGN from the ground-based Cherenkov telescopes are often triggered by $\gamma$-ray flares detected at MeV-GeV energies \citep[e.g.,][]{2015ApJ...815L..22A}. Moreover, a $\gamma$-ray spectral hardening during the elevated activity states has been reported for several VHE-detected blazars. Interestingly, the bright FSRQ PKS 1510$-$089 has also been detected in the VHE band during its quiescent state \citep[][]{2018A&A...619A.159M}. Therefore, it is crucial to investigate the source activity state and spectral behavior at the time of arrival of the VHE photon. We collected the monthly-binned 0.1$-$100 GeV light curve data of all $TS>25$ sources from the Fermi-LAT light curve repository \citep[][]{2023ApJS..265...31A}. We show the plots in Figure~\ref{lightcurves} where the VHE photon arrival times are highlighted with vertical dashed lines. We discuss observational results about individual sources in the next section. VHE photons were detected during periods of high-energy flaring activity of many sources, e.g., 4FGL J0348.5$-$2749; however, they were also detected during periods of low activity, e.g., 4FGL~J1031.1+7442.  Particularly, out of 47 VHE photons detected with $>$95\% source association probability, the arrival time of 24, i.e., 51\%, was coincided with flaring activity period. The remaining 23 photons, i.e., 49\%, were detected during period of quiescence. This suggests that VHE radiation may not be linked solely to flares but can also be emitted during low-activity periods, consistent with results found for high-synchrotron peaked BL Lac objects \citep[][]{2026ApJS..282...35A}.

\subsection{Location of the VHE-Emitting Region}

The BLR photon field can provide seed photons for inverse Compton scattering in FSRQs. The same BLR photons can also attenuate VHE photons via $\gamma\gamma$ pair production \citep[e.g.,][]{2003APh....18..377D,2010ApJ...717L.118P,2016ApJ...821..102B}. To quantify the optical depth for the $\gamma\gamma$ pair production ($\tau_{\gamma\gamma}$), we adopted the prescriptions of \citet{2016ApJ...830...94F} that have been coded in the package {\tt agnpy} \citep[][]{2022A&A...660A..18N}. We assumed a spherical BLR geometry and the BLR luminosity to be $10\%$ of accretion disk luminosity, following  \citet[][]{2021ApJS..253...46P}. We derived  $\tau_{\gamma\gamma}$ as a function of photon energy for different distances of the emission region from the Lyman-$\alpha$ photon field. We show the estimated opacity in Figure~\ref{opacity}, where vertical dotted lines represent the energies of the detected VHE photons. For all sources, we found that the $\gamma$-ray-emitting region must lie outside the BLR to allow the VHE radiation to escape. These results are similar to that found by \citet{2018MNRAS.477.4749C} for the general FSRQ population. The details of individual FSRQs are further discussed in section~\ref{individual}. 

\subsection{Cosmic Gamma-ray Horizon}

Beyond intrinsic jet energetics and spectral properties, the detectability of VHE $\gamma$-ray sources is critically influenced by attenuation from EBL, which restricts $\gamma$-ray propagation over cosmological distances {\bf \citep[e.g.,][]{1970Natur.226..135F}}. To examine this effect for the identified VHE emitting FSRQs, we plot the energy of the highest energy photon as a function of redshift in Figure~\ref{Cosmic horizon}. The cosmic $\gamma$-ray horizon, defined by an EBL optical depth of $\tau_{EBL}$=1, is also overlaid as a solid curve based on the model by \citet{2021MNRAS.507.5144S}. The majority of the VHE emitting FSRQs are located around this horizon, suggesting that they reside in regions where the Universe remains relatively transparent to VHE $\gamma$-rays. This highlights their suitability as detectable targets for both current and upcoming ground-based $\gamma$-ray facilities. Interestingly, one of the blazars, 4FGL~J2326.2+0113 (associated with SDSS~J232625.63+011208.6 at $z=1.6$) was found to have an EBL optical depth of $\tau_{EBL}\approx4$. Based on the {\tt gtsrcprob} analysis, a single VHE photon of 177.78 GeV was detected from this with a source association probability of $97\%$. This photon-pair conversion took place in the back-section of the LAT tracker (convtype=1), which has relatively poor angular reconstruction compared to events converted in the front conversion layer. Moreover, no other VHE photon was detected down to a source association probability of 50\%. Therefore, a strong claim regarding the VHE photon detection from this object cannot be made. However, if Fermi-LAT detects more VHE photons from this source, it would be a promising candidate to examine the current EBL models \citep[e.g.,][]{2015ApJ...813L..34D}.

\section{Individual notes on the VHE-emitting FSRQs (TS>25)}
\label{individual}

${\tt 4FGL~J0102.8+5824}$: This object is located at a redshift of 0.644 \citep[][]{2012ApJ...748...49S}. It is not present in TeVCat, thus, it is reported  for the first time  as a VHE emitter in this work. Its monthly binned $\gamma$-ray light curve revealed that, out of five VHE photons (association probability $>$95\%), no significant gamma-ray flaring activity was observed during four of the VHE detection epochs (Figure \ref{lightcurves}). However, during the epoch when a VHE photon was detected (MJD 60726-60756), a significant flare was also observed, accompanied by considerable spectral hardening (photon index $2.0\pm0.05$; Figure \ref{lightcurves}). These results hint that VHE radiation can be emitted during flaring as well as quiescent phases. The VHE spectrum of the source is steep (photon index = $5.2\pm2.3$), which is similar to that usually observed from VHE-detected broad emission line blazars \citep[e.g.,][]{2011ApJ...730L...8A}. From the estimated BLR optical depth variation (Figure \ref{opacity}), we found that, to avoid significant absorption by BLR photons, the VHE emission region must be located beyond $1.3\times$$R_{Lya}$, where $R_{Lya}$ represents the characteristic radius of the Lya-emitting region of the BLR \citep[cf.][for details]{2022A&A...660A..18N}. This constraint supports a scenario in which the emission region lies outside the dense photon field of the BLR, allowing VHE photons to escape without substantial attenuation.

$\tt{4FGL\ J0348.6-2749}$: This source is located at a redshift of 0.991, placing it at a considerable cosmological distance, where attenuation due to EBL is expected to significantly influence the observed $\gamma$-ray spectrum. This object was also detected with H.E.S.S. during a period of elevated \gm-ray activity \citep[][]{2026A&A...706A.246H}. Therefore, our work has confirmed its recognition as a VHE-emitting blazar. Three VHE photons were detected by the Fermi-LAT with energies of 101.94 GeV, 102.08 GeV and 103.55 GeV, all during \gm-ray flaring states (Figure \ref{lightcurves}). During the VHE emission period, the 0.1$-$100 GeV \gm-ray spectrum remained hard (photon index $\sim2$). The VHE spectrum is steep, with a photon index of $17.8\pm0.8$, and could not be constrained after applying EBL correction, likely due to poor photon statistics (Table~\ref{tab:source details}). It is the faintest VHE-emitting object in our sample. From the plot of optical depth as a function of energy (Figure \ref{opacity}), we found that, to avoid significant absorption within the BLR, the VHE emission region must be located beyond $1.4\times$$R_{Lya}$.

$\tt{4FGL\ J0428.6-3756}$: This source is located at a redshift of 1.11, and it is the most distant VHE-detected object in our sample. Previously, \citet[][]{2013ApJ...777L..18T} reported the detection of two VHE photons from this object. Significant $\gamma$-ray flaring activity was observed during the emission of two VHE photons \citep[see also;][]{2013ApJ...777L..18T}. Interestingly, the third VHE photon was detected during a quiescent period (Figure \ref{lightcurves}). A hardening of the 0.1$-$100 GeV \gm-ray spectrum was noticed during the arrival of the VHE photons (Figure \ref{lightcurves}). From the evaluation of optical depth as a function of photon energy for different locations of the emission region relative to the Lyman-$\alpha$ BLR photon field (Figure \ref{opacity}), we inferred that the VHE emission region must be beyond $1.2\times$$R_{Lya}$. We note that this \gm-ray source, associated with PKS 0426$-$380, was identified as a transitional BL Lac object with broad emission lines seen in the optical spectrum taken during a low jet activity state \citep[][]{2005AJ....129..559S,2011MNRAS.414.2674G}.

$\tt{4FGL\ J0538.8-4405}$: This \gm-ray blazar, located at a redshift of 0.896, is associated with the radio source PKS 0537$-$44. This object is also a transitional BL Lac object with broad optical emission lines detected during low jet activity \citep[][]{2002A&A...392..407P}. This source is not listed in TeVCat. Four VHE photons, with energies 104.32 GeV, 121.12 GeV, 125.93 GeV and 145.72 GeV, were detected, supporting its classification as a promising VHE-emitting source. The 0.1$-$100 GeV light curve revealed that three VHE photons arrived during its high brightness state, while one photon was detected during a relatively quiescent phase (MJD 57895; Figure \ref{lightcurves}). The VHE spectrum of the source was steep (photon index = $5.7\pm2.4$). After applying the EBL attenuation correction, the intrinsic spectrum becomes comparatively harder with a photon index of $3.2\pm2.7$ (Table~\ref{tab:source details}). From the optical depth profile as a function of energy (Figure \ref{opacity}), we inferred that the VHE emission region must be located beyond $1.4\times$$R_{Lya}$, supporting a scenario in which VHE photons are produced outside the dense photon field of the BLR, allowing them to escape without substantial attenuation.

$\tt{4FGL\ J0904.9-5734}$: The \gm-ray source was recently detected in the VHE band by H.E.S.S. during a period of high activity detected with the Fermi-LAT \citep[][]{2020ATel13632....1W}. \citet[][]{2024A&A...691L...5G} reported its redshift to be $z=0.262$, which is different from $z=0.697$ adopted in previous works. In our work we have adopted $z=0.262$. This object is the most significantly detected VHE source in our sample (TS=82, Table~\ref{tab:source details}). The Fermi-LAT detected nine VHE photons ($>$95\% association probability), including a VHE photon of $\sim$480 GeV. The arrival times of the VHE photon coincided with flaring as well as quiescent activities, as revealed by its monthly-binned \gm-ray light curve (Figure \ref{lightcurves}). A spectral hardening was noticed in the 0.1$-$100 GeV spectrum at the time of arrival of most of the VHE photons. Estimation of optical depth associated with  \gm\gm~pair production, shown in Figure \ref{opacity}, revealed that the VHE emission region must be located beyond $1.1\times$$R_{Lya}$.

$\tt{4FGL\ J0957.6+5523}$: This \gm-ray blazar is associated with the radio source 4C +55.17 ($z=0.903$). It was proposed as a promising VHE emitting FSRQ due to its hard 0.1$-$100 GeV spectrum and lack of flux variability. It was observed with MAGIC telescopes for 50 hours but remained undetected \citep[][]{2014MNRAS.440..530A}. Therefore, this work is the first report of the VHE detection of this FSRQ at a confidence level greater than $5\sigma$. Its monthly binned 0.1$-$100 GeV light curve does not reveal any significant flux variations, thus all VHE photons arrived during the \gm-ray low-activity periods (Figure \ref{lightcurves}). The measured 0.1$-$100 GeV photon index remained less than 2, indicating consistently hard spectral states (Figure \ref{lightcurves}). In contrast, the observed VHE spectrum is steep (photon index $4.8\pm2.1$). Interestingly, the intrinsic VHE spectrum, after EBL attenuation correction, is considerably harder, though the photon index uncertainty is large, which precluded us from making a strong claim. From the optical depth calculation, it is inferred that the location of the VHE emitting region has to be beyond $1.4\times$$R_{Lya}$.

$\tt{4FGL\ J1031.1+7442}$: This \gm-ray source, at $z=0.123$, was recently reported as a VHE emitter by \citet{2025ApJ...991L...8P}. As also reported by them, all VHE photons were detected during quiescent periods (Figure \ref{lightcurves}). This suggests that VHE emission can originate from steady or mildly variable jet conditions rather than being exclusively flare-driven as usually detected for other VHE detected \mbox{FSRQs}. The VHE spectrum of the source is relatively hard (photon index = $2.2\pm0.8$) compared to other VHE-emitting \mbox{FSRQs} (Table~\ref{tab:source details}). Optical depth considerations constrain the location of the VHE emission region to be outside BLR, i.e., greater than $R_{Lya}$ (Figure \ref{opacity}).

$\tt{4FGL\ J1224.9+2122}$: This \gm-ray bright blazar, associated with 4C +21.35 ($z=0.434$), was first identified as a VHE emitting FSRQ by MAGIC during a \gm-ray outburst in 2011 \citep[][]{2011ApJ...730L...8A}. As per our analysis, several VHE photons were detected from this object with the highest energy photon of 256.55 GeV (Table~\ref{tab:all sources}). Temporal analysis (Figure \ref{lightcurves}) indicated that most of the VHE photons were detected during \gm-ray flaring episodes, while one VHE photon was observed during a quiescent state. Similar to other VHE-emitting FSRQs, the 0.1$-$1 TeV spectrum is soft (photon index = $3.1\pm1.2$). The EBL corrected spectrum, on the other hand, is harder, with a photon index of $1.6\pm1.4$. The BLR opacity calculation indicates that the VHE-emitting region is likely to be located beyond $1.4\times$$R_{Lya}$ to avoid absorption of the VHE photons by the dense BLR photon field.

$\tt{4FGL\ J1310.5+3221}$: At a redshift of 0.996, this object (associated with OP 313) is one of the most distant FSRQs detected in the VHE band by the Cherenkov telescopes \citep[][]{2023ATel16381....1C,2025ATel17000....1P}. The Fermi-LAT detected five VHE photons with association probability greater than 95\%. The 0.1$-$100 GeV light curve revealed that all VHE photons were detected during periods of \gm-ray outbursts, indicating that enhanced activity plays a role in driving VHE emission in this source (Figure ~\ref{lightcurves}). The corresponding MeV$-$GeV spectrum was hard during all epochs of VHE photon detection (0.1$-$100 GeV photon index less than 2). The VHE spectrum, on the other hand, is steep with a 0.1$-$1 TeV photon index of \mbox{$8.9.\pm4.5$} Unlike other sources, the EBL attenuation corrections did not lead to an intrinsic harder spectrum, likely due to limited photon statistics. Furthermore, to avoid severe $\gamma-\gamma$ absorption within BLR, the VHE emission site has to be located beyond $1.3\times$$R_{Lya}$, implying that the emission region lies outside the dense BLR photon field.

$\tt{4FGL\ J1512.8-0906}$: This bright \gm-ray source is associated with the well-known radio source PKS 1510$-$089 at a redshift of 0.362. It is the only known VHE emitting FSRQ that has been detected by Cherenkov telescopes during low activity periods \citep[][]{2018A&A...619A.159M}. Several VHE photons were detected with an association probability greater than 95\%, yielding a statistically significant detection (Table~\ref{tab:source details}). Scanning the 0.1$-$100 GeV light curve, we found that VHE photons arrived both during flaring as well as low activity states. The VHE spectrum of the source is soft (photon index = $3.2\pm0.9$), which becomes harder after applying EBL-correction (photon index = $2.0\pm1.1$). Constraints from optical depth estimation indicate that, to avoid significant $\gamma-\gamma$ absorption within BLR, the emission region must be located beyond $1.2\times$$R_{Lya}$, implying that VHE emission originates outside the BLR, enabling photons to escape with minimal internal attenuation (Figure \ref{opacity}).

\label{subsec:source_stats}
\section{Summary}
\label{sec:summary}
This study investigates VHE $\gamma$-ray emission from a sample of Compton-dominated blazars using $\sim$17.5 years of the Fermi-LAT data. From an initial sample of 626 objects, 10 sources are significantly detected in the VHE band (TS $>$ 25), including 4 for the first time. We also found 4 sources with tentative VHE emission (16$<$TS$<$25), and 21 additional candidate sources exhibiting at least one VHE photon. We investigated the connection of the VHE photon detection with the \gm-ray activity, and found that VHE emission can occur during both flaring and quiescent states. The observed VHE spectra are generally steep but become harder after correcting for attenuation by EBL. By modelling $\gamma-\gamma$ absorption within BLR, the study constrains the emission region to be outside the BLR, typically beyond 1.1-1.4 $\times$ $R_{Lya}$. Additionally, the distribution of highest-energy photons with redshift suggests that most detected sources lie within or near the cosmic $\gamma$-ray horizon.

Overall, the work significantly expands the known population of VHE-emitting FSRQs and provides important insights into their emission mechanisms, jet environment, and $\gamma$-ray production regions. These objects may be considered as the prime targets for deeper VHE observations with the Cherenkov Telescope Array Observatory, thanks to its excellent sensitivity and broad energy coverage.

\section*{Acknowledgements}
We thank the journal referee for constructive criticism and comments which have helped improve the manuscript. This research has made use of NASA's Astrophysics Data System Bibliographic Services. The use of the \textit{Fermi}-LAT data provided by the Fermi Science Support Center is gratefully acknowledged. The author sincerely appreciates the research facilities provided by UGC-SAP and FIST 2 (SR/FIST/PS1-159/2010) (DST, Government of India) in the Department of Physics, University of Calicut. We also extend our gratitude to DST-FIST for providing research facilities at Farook College (Autonomous), Calicut.

\section*{Data Availability}
The data utilized in this study are publicly accessible and were obtained from the archives available at \url{https://Fermi.gsfc.nasa.gov/}.


\appendix
\clearpage
\bibliographystyle{elsarticle-harv} 
\bibliography{VHE}
\end{document}